\documentclass[journal]{IEEEtran}

\ifCLASSINFOpdf
\else
   \usepackage[dvips]{graphicx}
\fi

\usepackage{color}
\usepackage[ruled]{algorithm2e}
\usepackage{graphicx}
\usepackage{amsmath}
\usepackage{algorithmic}
\usepackage{array}
\usepackage{verbatim} 
\usepackage[english]{babel}
\DeclareMathOperator*{\argmin}{arg\,min}

\graphicspath{{./}}

\begin{document}
%
\title{Content-Preserving Image Stitching with Regular Boundary Constraints}
%
%
%

\author{Yun Zhang,
        Yu-Kun Lai,~\IEEEmembership{Member,~IEEE,}
        and~Fang-Lue Zhang,~\IEEEmembership{Member,~IEEE}
\thanks{Yun Zhang is with the Institute of Zhejiang Radio and TV Technology, Communication University of Zhejiang, Hangzhou, China, 310018.\protect\\
E-mail: zhangyun@cuz.edu.cn}
\thanks{Yu-Kun Lai is with the School of Computer Science and Informatics, Cardiff University, Wales, UK,  CF24 3AA.\protect\\
E-mail: Yukun.Lai@cs.cardiff.ac.uk}
\thanks{Fang-Lue Zhang is with the School of Engineering and Computer Science, Victoria University of Wellington, New Zealand.\protect\\
E-mail: fanglue.zhang@ecs.vuw.ac.nz}}

\markboth{Submitted to IEEE TVCG}%
{Shell \MakeLowercase{\textit{et al.}}: Bare Demo of IEEEtran.cls for IEEE Journals}

\maketitle

\begin{abstract}
This paper proposes an approach to content-preserving image stitching with regular boundary constraints, which aims to stitch multiple images to generate a panoramic image with regular boundary. Existing methods treat image stitching and rectangling as two separate steps, which may result in suboptimal results as the stitching process is not aware of the further warping needs for rectangling. We address these limitations by formulating image stitching with regular boundaries in a unified optimization. Starting from the initial stitching results produced by traditional warping-based optimization, we obtain the irregular boundary from the warped meshes by polygon Boolean operations which robustly handle arbitrary mesh compositions, and by analyzing the irregular boundary construct a piecewise rectangular boundary. Based on this, we further incorporate straight line preserving and regular boundary constraints into the image stitching framework, and conduct iterative optimization to obtain an optimal piecewise rectangular boundary, thus can make the  boundary of stitching results as close as possible to a rectangle, while reducing unwanted distortions. We further extend our method to selfie expansion and video stitching, by integrating the portrait preservation and temporal coherence into the optimization. Experiments show that our method efficiently produces visually pleasing panoramas with regular boundaries and unnoticeable distortions.

\end{abstract}


\IEEEpeerreviewmaketitle

\section{Introduction}

The rapid recent advances in digital visual media mean that the public can now capture and produce high-quality images and videos, which has promoted the computer graphics applications that utilize visual data captured by ordinary users. Image/video panorama  is one of these successful applications. With the integrated panorama module in their smart phones and portable cameras, people can easily take panoramic photos by simply moving their cameras. It is also the most accessible way to get virtual reality content for immersive visual experience. However, unlike well calibrated images captured by professional devices with a camera array, the intrinsic and extrinsic parameters of the images captured by consumer-level devices are difficult to estimate. Thus, robust image stitching methods which directly stitch visual content is highly important for such applications designed for ordinary users.

\begin{figure*}[t] 
  \centering
  \includegraphics[width=1.0\textwidth]{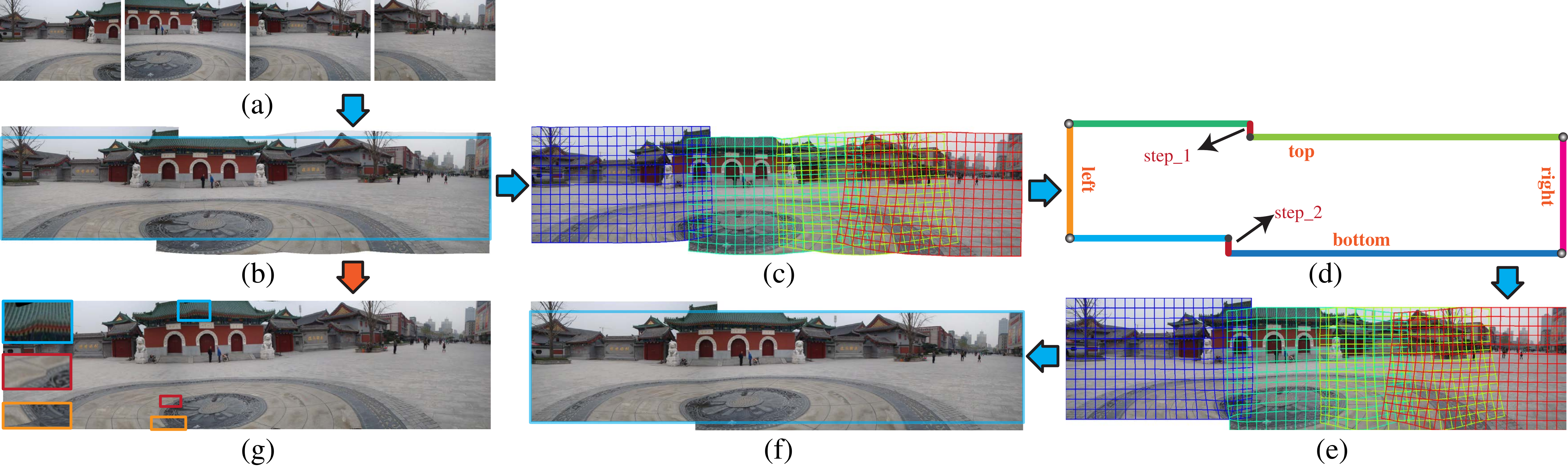}
  \caption{Pipeline of our image stitching method with regular boundaries: (a) input images, (b) initial stitching with an irregular boundary, (c) meshes of initial stitching, (d) piecewise rectangular boundary, (e) warped meshes for piecewise rectangling, (f) our result, (g) rectangular panorama result by~\cite{journals/tog/HeC013} .}
  \label{fig:pipeline}
\end{figure*}

Recently,  much progress has been made in image stitching. However, due to the casual motion of hand-held cameras, most stitching results by existing methods have irregular boundaries after the local feature alignment. But the common application scenario for stitched images is to display the full panorama on normal screens, or generate free-viewpoint photos from part of the whole scene recorded as an image collection, which means that we can only show them in rectangular windows.
To achieve this, a simple and direct method is cropping, but it usually causes loss of important content in the stitched panorama, and reduces the impression of wide angle photography.
In order to produce panoramic images with rectangular boundaries, the image completion technique~\cite{journals/mta/YenYC17, journals/tog/BarnesSFG09} is used to synthesize missing regions in the bounding box of panoramic images.
However, these methods are not stable, and may fail when synthesizing regions with rich structures and semantically meaningful objects.

He et al.~\cite{journals/tog/HeC013} proposed ``\emph{rectangling}'' to produce visually pleasing panoramic images with desired rectangular boundaries by warping the initial stitched panoramas.
Although effective in many examples, their method suffers from the following problems:
(1) The stitching and rectangling are two separate processes, so the latter rectangling step may distort the optimized stitching result, making it hard to get an optimal rectangular panorama. Moreover, making arbitrary boundaries rectangular may also introduce excessive distortions unacceptable for target applications.
(2) Their method relies on placing a grid mesh on a stitched irregular panorama for rectangling, where the grid may contain pixels out of the stitched image due to the boundary irregularity, leading to the resulting rectangular image with small holes on the boundary (see Fig.~\ref{fig:holes_he}(d)).
(3) The warping-based method may cause large distortions and destroy feature alignment, when turning an incompletely shot scene to a rectangle.
In summary, when the gap between the rectangular boundary and irregular panorama boundary is large, or there are holes which are difficult to fill in by inpainting or warping,
a better approach is demanded to create panoramic images with regular boundaries while avoiding large distortions.

In this paper, we propose a novel approach for content-preserving image stitching, which aims to regularize the boundary of the stitched panorama, and preserves as much content as possible in a rectangular cropping window.
Our method is based on the following observations:
(1) Rectangling and stitching are tightly related, and optimizing the  two processes simultaneously can help produce better rectangular panoramas in a content-aware manner.
(2) The aim of panorama rectangling is to preserve as much  image content as possible in a rectangular window while avoiding unexpected distortions. To achieve this,
an irregular boundary should not be simply optimized to be a single rectangle as evidenced by Fig.~\ref{fig:pipeline}(g). We propose to instead use a more flexible piecewise rectangular boundary (see Fig.~\ref{fig:pipeline}(d) for an example) to ensure regularity while avoiding excessive distortions. Using piecewise rectangular boundaries also has the advantage that treats traditional rectangular boundaries as a special case, and will provide rectangular results when appropriate.

Our method works well even for challenging cases and can produce visually pleasing results without user interactions (see Fig.~\ref{fig:challenging_cases} for some examples).
After stitching input images using a traditional method with the global similarity prior~\cite{conf/eccv/ChenC16}, we extract the outer boundary of the stitching result and analyze the boundary constraints, and finally perform a global optimization taking these constraints into account to obtain the stitching result with a piecewise rectangular boundary.
Our method can robustly stitch a large number of images. To achieve this, we treat each image in the initial stitching result as a warped mesh, and utilize polygon Boolean operations to extract irregular boundaries and suitable boundary constraints for piecewisely rectangling. In the global optimization stage, we take into account the regular boundary, shape preserving, straight lines and global similarity constraints in a unified optimization framework.
To obtain panoramic images with optimal piecewise rectangular boundaries, we firstly automatically extract a piecewise rectangular boundary (see Fig. 1(d)), then iteratively combine boundary segments connected by steps to simplify the shape of panorama boundary while avoiding large distortions. 
Finally, after minimizing the energy function, we get the stitching result by warping and blending.
When the target boundary is simply a rectangle, our method performs stitching and rectangling simultaneously, and can produce panoramas with a rectangular boundary (see some examples in Fig.~\ref{fig:more_results}).
Our method can help users easily crop panoramas while preserving as much content as possible in the cropping window, and avoiding unwanted distortions, thus can enhance the panorama viewing experiences.
Furthermore,  our method can be extended to obtain panoramic selfies and videos with regular boundaries.

The main contributions of this paper are summarized as follows:
\begin{itemize}

   \item We propose a global optimization approach to producing panoramic images by simultaneously stitching images and optimizing boundary regularity in a unified framework. By doing so, our method reduces undesired distortions compared with traditional approaches where stitching and rectangling are treated as two separate steps.


   \item We propose to use piecewise rectangular boundaries to achieve regular boundaries while preserving content from input images as much as possible and avoiding excessive distortions compared with traditional rectangling, and further develop a fully automatic algorithm to produce optimized piecewise rectangular boundaries to balance the distortion and boundary simplicity.
\end{itemize}


 \section{RELATED WORK}
In this section, we briefly review the techniques most related to our work.

\textbf{Image stitching}.
Image stitching aims to create seamless and natural photo-mosaics. A comprehensive survey of image stitching algorithms is given in~\cite{journals/ftcgv/Szeliski06}.
Brown et al.~\cite{journals/ijcv/BrownL07} proposed a method for fully automatic panoramic image stitching, which aligns multiple images by a single homography. Their method is effective under the assumption that the camera only rotates around its optical center, the images are shot from the same viewpoint and the scenes are nearly planar. However, for images shot by hand-held cameras, they always contain parallax, which limits the application of their method.
Given the limitation of single homography, Gao et al.~\cite{conf/CVPR/GaoKB11} proposed to use two homographies to perform nonlinear alignment, where the scene is modeled by dominant distant and ground planes. However, their method is only effective when there are no local perspective variations.

For better performance in image alignment, Zaragoza et al.~\cite{journals/pami/ZaragozaCTBS14} proposed as-projective-as-possible (APAP) warping based on the Moving Direct Linear Transformation (DLT), and can seamlessly align images with different projective models. Their method can handle global perspectives, while allowing local non-projective deviations, thus can deal with some challenging cases.
This technique has been widely applied in image alignment due to its excellent performance and in this paper we also use APAP for our initial stitching before optimization.
Based on APAP, researchers attempted to get more natural panoramas. Lin et al.~\cite{conf/CVPR/LinPRA15} combined local homography and global similarity transformation to achieve more continuous and smooth stitching results. It provides stitched panoramas with less visible parallax and perspective distortions.
Li et al.~\cite{conf/ICCV/LiY0Q15} proposed a dual-feature warping-based model by combining keypoints and line segment features. However, the 2D model proposed in this paper cannot handle large parallax and depth variations, and it is difficult to determine the line correspondences in images with large parallax.
Chang et al.~\cite{conf/cvpr/ChangSC14} proposed a parametric warping method which combines projective and similarity transformation. By combining APAP~\cite{journals/pami/ZaragozaCTBS14}, their method can significantly reduce  distortions in stitching results.
Chen et al.~\cite{conf/eccv/ChenC16} further proposed natural image stitching with global similarity prior. They designed a selection scheme to automatically determine the proper global scale and rotation for each image.

There are also methods focusing on local alignment adjustment for eliminating stitching artifacts.
To stitch images with large parallax, Zhang et al.~\cite{conf/cvpr/ZhangL14a} proposed a local stitching method, which is based on the observation that overlapping regions do not need to be aligned perfectly. Lin et al.~\cite{conf/eccv/LinJCDL16} proposed a seam-guided local alignment approach where optimal local alignment is guided by the seam estimation.
In their method, salient curves and line structures are preserved by local and non-local similarity constraints.
Very recently, Li et al.~\cite{journals/tmm/LiWLZZ18} proposed robust elastic warping for parallax-tolerant image stitching. To ensure robust alignment, they proposed a Bayesian model to remove incorrect local matches.

However, none of  these methods above consider how to achieve better results in the display window. He et al.~\cite{journals/tog/HeC013} proposed a content-aware warping method to produce rectangular images from the stitched panoramas.  Their method is effective to rectify irregular boundaries caused by projections and casual camera movements. However, their two-step warping strategy separates the stitching and rectangling processes, and therefore cannot ensure optimal solutions. Moreover, their method cannot cope well with scenes that are not completely captured.
Unlike~\cite{journals/tog/HeC013}, we incorporate stitching and rectangling into a unified framework, and construct global optimization to obtain piecewise rectangular panoramic images.

\textbf{Video stitching}.
Compared with image stitching, video stitching is  more difficult, due to the camera motion, dynamic foreground and large parallax.
For static camera settings, such as multi-camera surveillance~\cite{journals/sensors/HeY16,journals/itiis/YinLWLZ14},  videos from different cameras are aligned only once, and the main challenge is to avoid ghosting and artifacts caused by moving objects.
For moving cameras with relatively fixed positions, such as a camera array fixed on a rig~\cite{journals/cgf/PerazziSZKWWG15}, cameras can be pre-calibrated for global stitching of videos, and spatio-temporally coherent warping and minimizing distortion are the main challenges due to the motion and parallax.
Google Street View~\cite{journals/computer/AnguelovDFFLLOVW10} also utilized moving camera arrays for street view capture and panorama generation.
To generate high-quality panoramic videos for those captured by a camera array fixed on a rig, Zhu et al.~\cite{journals/tip/ZhuLWZMLH18} proposed a method for real-time panoramic video blending.
Meng et al.~\cite{conf/mm/MengWL15} proposed a multi-UAV (unmanned aerial vehicle) surveillance system that supports real-time video stitching.
Recently, many researchers focused on stitching algorithms for videos shot by multiple hand-held cameras.
El-Saban et al.~\cite{conf/icip/El-SabanEKR11} proposed optimal seam selection blending for fast video stitching; however, they do not consider video stabilization.
Lin et al.~\cite{journals/cgf/LinLCZ16} firstly proposed a robust framework to stitch videos from moving hand-held cameras, which incorporates stabilization and stitching into a unified framework.
Guo et al.~\cite{journals/tip/GuoLHZZG16} and Nei et al.~\cite{journals/tip/NieSZSL18} further improved the performance of  a joint video stabilization and stitching framework.
Their main contributions include: estimation of inter-motions between cameras and intra-motions in a video, and common background identification for multiple input videos.
In this paper, we further extend our content-preserving image stitching to videos that are captured from unstructured camera arrays~\cite{journals/cgf/PerazziSZKWWG15}.

 \section{Overview}
Fig.~\ref{fig:pipeline} gives the pipeline of our content-preserving stitching method.
The input to our approach is a number of images with partial overlaps, and the goal is to obtain a panoramic image with regular boundary.
Similar to previous warping-based stitching, we place separate quad mesh on each image, and construct energy functions with constraints on the image meshes.
The core of our approach is a unified optimization framework that combines image stitching and piecewise rectangling, which contains the following key steps:

\textbf{Preprocessing}.
We first calculate the image match graph using the method proposed in~\cite{journals/ijcv/BrownL07}. The images that are connected in the match graph are aligned in the stitching process. This automatic matching process allows stitching with complex image overlaps (see Fig.~\ref{fig:piecewise-compare}(b)).
For straight line and global feature preserving, we detect lines in all images using the fast line segment detector~\cite{journals/pami/GioiJMR10}.

\textbf{Initial image stitching}.
The goal of this step is to initialize our content-preserving stitching, which also provides the basis for analyzing regular boundary constraints.
The stitching strategy in this step is also incorporated into the optimization of our  piecewise rectangling stitching.
We apply APAP~\cite{journals/pami/ZaragozaCTBS14} for accurate feature alignment.
Inspired by~\cite{conf/eccv/ChenC16}, we also add a global similarity term for more natural stitching with less distortion.

\textbf{Piecewise rectangular stitching}.
After the initial image stitching, we extract the contour of each warped mesh, and obtain the irregular boundary of the stitching result by polygon Boolean union operations.
Then, we analyze vertices and intersections on the irregular boundary to get regular boundary constraints for our energy optimization, and
iteratively optimize the piecewise rectangular boundary by combining boundary segments connected by each step on the regular boundary, to achieve a good balance between the boundary simplicity and content distortions.
Finally, we minimize the energy function, and get the stitching result by warping and blending.

\section{Initial image stitching}
\label{sec:Initial image stitching}
%
After preprocessing, we propose to stitch images using a content-preserving approach. The warped meshes of the stitched images with irregular boundaries will serve as an initial values for the optimization to get piecewise rectangular stitching result.
Inspired by~\cite{conf/eccv/ChenC16}, we stitch images using the global similarity prior to generate more natural panoramas without too much distortion or limited field of view.
Like previous methods, image stitching is performed by mesh-based image warping on input images. Each input image is represented using a regular quad mesh placed on it.
Let $V = \{V^i\}$ and $E = \{E^i\}$ be the sets containing all the vertices and edges of input images, where $i=1, 2, \dots, N$, and $N$ is the number of images to be stitched.
The $j^{\rm th}$ vertex of $V^i$ is then represented as $V^i_j$. For simplicity, we also use $V^i_j$ to represent the vertex position without ambiguity.
 We aim to obtain the deformed vertices $\hat{V}$ by minimizing the energy function $\Phi(\hat{V})$,
which contains the following terms: feature alignment, local shape preserving and global similarity preserving.

\textbf{Feature Alignment}.
This term aims to align matched images by preserving their feature correspondences after deformation.
Given its good performance in piecewise alignment, we apply APAP~\cite{journals/pami/ZaragozaCTBS14} for feature alignment, which is defined as follows based on each matched pair in the match graph of all images from the preprocessing step.
\begin{equation} \label{equ:feature_align}
\phi_a(\hat{V})=
\sum\limits_{(i, j)\in G}\sum\limits_{m_k^{ij} \in M^{ij}}\|\tilde{v}(m_k^{ij}(i))-\tilde{v}(m_k^{ij}(j)) \|^2,
\end{equation}
where $G$ refers to the image match graph which contains all the matched image pairs $(i, j)$.
$m^{ij}_k$ represents a pair of matched feature points from images $i$ and $j$, and $M^{ij}$ is the set of all the feature matchings for image pair $(i,j)$.
Since the constraints are imposed on mesh vertices and the matched points are in general not mesh vertices, we employ the bilinear coordinate representation for the matched feature points.
In Equ.~\ref{equ:feature_align}, $m_k^{ij}(i)$ and $m_k^{ij}(j)$ refer to the feature points on images $i$ and $j$ respectively for the matched feature pair $m^{ij}_k$.
Denote by $\tilde{v}(m_k^{ij}(i))$ the position of the deformed feature point, which is represented
 by interpolating the vertex positions of the mesh grid on image $i$ that contains $m_k^{ij}(i)$.
Specifically, $\tilde{v}(m_k^{ij}(i)) = \mathbf{\hat{V}}^{i}_{pq} \cdot  \boldsymbol{\Omega}^{i}_{pq}$ , where $\cdot$ is the dot product, $p$ and $q$ are the indexes of the quad,  $\mathbf{\hat{V}}^{i}_{pq}=[\hat{V}^{i}_{p,q}, \hat{V}^{i}_{p+1,q}, \hat{V}^{i}_{p+1,q+1}, \hat{V}^{i}_{p,q+1}]$ are the positions of the deformed mesh vertices, and $\boldsymbol{\Omega}^{i}_{pq}=[\omega^{i}_{p,q},\omega^{i}_{p+1,q},\omega^{i}_{p+1,q+1},\omega^{i}_{p,q+1}]$ are the interpolation weights that sum to 1, calculated based on the position of the feature point w.r.t. the grid vertices before warping. $\tilde{v}(m^{ij}_k(j))$ is similarly defined.
This term penalizes deviations of matched feature pairs after warping.

\textbf{Shape consistency}.
This term aims to ensure that grid quads in the image mesh undergo similar transforms and do not distort too much.
We use the shape preserving term defined in~\cite{journals/tog/LiuYT013}, which splits each grid quad into two triangles and applies as-rigid-as-possible warping~\cite{journals/tog/IgarashiMH05}.

\begin{equation} \label{equ:shape_preserving}
\begin{split}
    \phi_s(\hat{V}) = \sum\limits_{i=1}^N\sum\limits_{\hat{V}^i_j \in \hat{V}^i} \
    \|\hat{V}_{j}^i -\hat{V}_{j_1}^i-\xi \mathbf{R}(\hat{V}_{j_0}^i -\hat{V}_{j_1}^i)||^2\\
   \textrm{where}\quad \mathbf{R}=\left[\begin{array}{cc}\cos\theta&\sin\theta \\-\sin\theta&\cos\theta\end{array}\right]\\
    \textrm{and} \quad\xi=\|V_{j}^i -V_{j_1}^i\|/\|V_{j_0}^i -V_{j_1}^i\|,
\end{split}
\end{equation}
where $\xi$ and $\theta$ are the scaling and rotation parameters between vectors $\overline{V^i_{j_0} V^i_{j_1}}$ and $\overline{V^i_j V^i_{j_1}}$ in the initial mesh.
$\theta = \angle V_j^i V_{j_1}^i V_{j_0}^i$, and $V_j^i$, $V_{j_0}^i$, $V_{j_1}^i$ are neighboring vertices of each split triangle.
Since the original mesh grids are rectangles, the split triangles should preserve the \emph{right} angle, thus we set $\theta=90^{\circ}$.
 To obtain a  shape-preserving warping, the deformed vertices $\hat{V}_j^i$, $\hat{V}_{j_0}^i$, $\hat{V}_{j_1}^i$ should satisfy the similarity transform.

\textbf{Global similarity}.
We use the global similarity term proposed in~\cite{conf/eccv/ChenC16}, which is important to preserve the naturalness of panoramic images.
We first set image $I_1$ as reference, and specify its desired rotation angle $\theta_1$ with its scaling $s_1$ set to $1$. For any other image $I_i$ ($2 \leq i \leq N$), the desired scaling $s_i$ and rotation angle $\theta_i$ w.r.t. $I_1$ are calculated according to~\cite{conf/eccv/ChenC16}. The global similarity term is defined as
\begin{equation} \label{equ:global_similarity}
\begin{split}
\centering
    \phi_g(\hat{V}) = \sum\limits_{i=2}^N
    \sum\limits_{\hat{e}^i_j \in \hat{E}^i} \beta(\hat{e}^i_j)[&\| c_x(\hat{e}^i_j)-s_i \cos\theta_i \|^2+\\    &\| c_y(\hat{e}^i_j)-s_i \sin\theta_i \|^2],
\end{split}
\end{equation}
where $c_x(\hat{e}^i_j)$ and $c_y(\hat{e}^i_j)$ refer to the coefficients of  grid edges for similarity transforms in $x$ and $y$ directions; see details in~\cite{journals/jgtools/IgarashiI09}. $\hat{e}^i_j$ is a warped edge, which is determined by the warped vertices of the edge endpoints.
$\beta(\hat{e}^i_j)$ is used to assign more importance to edges in overlapping regions, while less in other regions, in order to keep accurate alignment and preserve naturalness, and it is defined as
\begin{equation} \label{equ:global_weight}
    \beta(\hat{e}^i_j)= \frac{\sum\limits_{q_m\in Q(\hat{e}^i_j)} d_{c}(q_m, \Psi_i)}{\sqrt {W_i^2+H_i^2}\cdot |Q(\hat{e}^i_j)|},
\end{equation}
where $|Q(\hat{e}^i_j)|$ refers to the number of quads that contains edge $\hat{e}^i_j$,
$\Psi_i$ is the region in image $I_i$  overlapping with other images,
$d_{c(q_m, \Psi_i)}$ calculates the minimum distance between the center of quad $q_m$ to quads in $\Psi_i$,
and $W_i$ and $H_i$ are the numbers of rows and columns of the mesh in image $I_i$.

With the energy terms above, we define the overall energy for image stitching as
\begin{equation} \label{equ:global_stitching}
 \Phi_{stitch}(\hat{V})=
\gamma_a \phi_a(\hat{V})+\gamma_s \phi_s(\hat{V})+\gamma_g \phi_g(\hat{V}),
\end{equation}
where $\gamma_a$, $\gamma_s$, $\gamma_g$ are used to control the importance of the three energy terms.
In our experiments, we fix $\gamma_a=1$ and set $\gamma_s=6.5$, $\gamma_g=0.5$ by default.
We give more importance to preserve the shape of image meshes for less distortions in the warping based optimization.

\section{Piecewise rectangular stitching}
For a given image collection, directly warping them to align with a single rectangle may not be preferable when large regions are missing.
For example, as shown in Fig.~\ref{fig:pipeline}(g), warping-based rectangling~\cite{journals/tog/HeC013} may introduce unwanted distortions when the gap between the irregular boundary of the initial stitching result and the target rectangular boundary is too large.
To avoid such undesirable artifacts, we propose to generate \emph{piecewise rectangular} boundaries which can make the target boundary as rectangular as possible, while avoiding excessive distortions if there are large missing regions in the whole scene. We also consider content-preserving constraints simultaneously when optimizing the warped meshes.
Compared with~\cite{journals/tog/HeC013}, using the piecewise rectangular boundary, the stitching result can be easily cropped into a rectangular photo, to display more content in a screen; see examples in Fig.~\ref{fig:challenging_cases}.
We firstly extract and analyze the irregular boundaries from the initial stitching results in Section~\ref{sec:Initial image stitching}, and then design the optimization objective for stitching that considers the piecewise rectangular boundary constraints.

\begin{figure}[t] 
  \centering
  \includegraphics[width=0.39\textwidth]{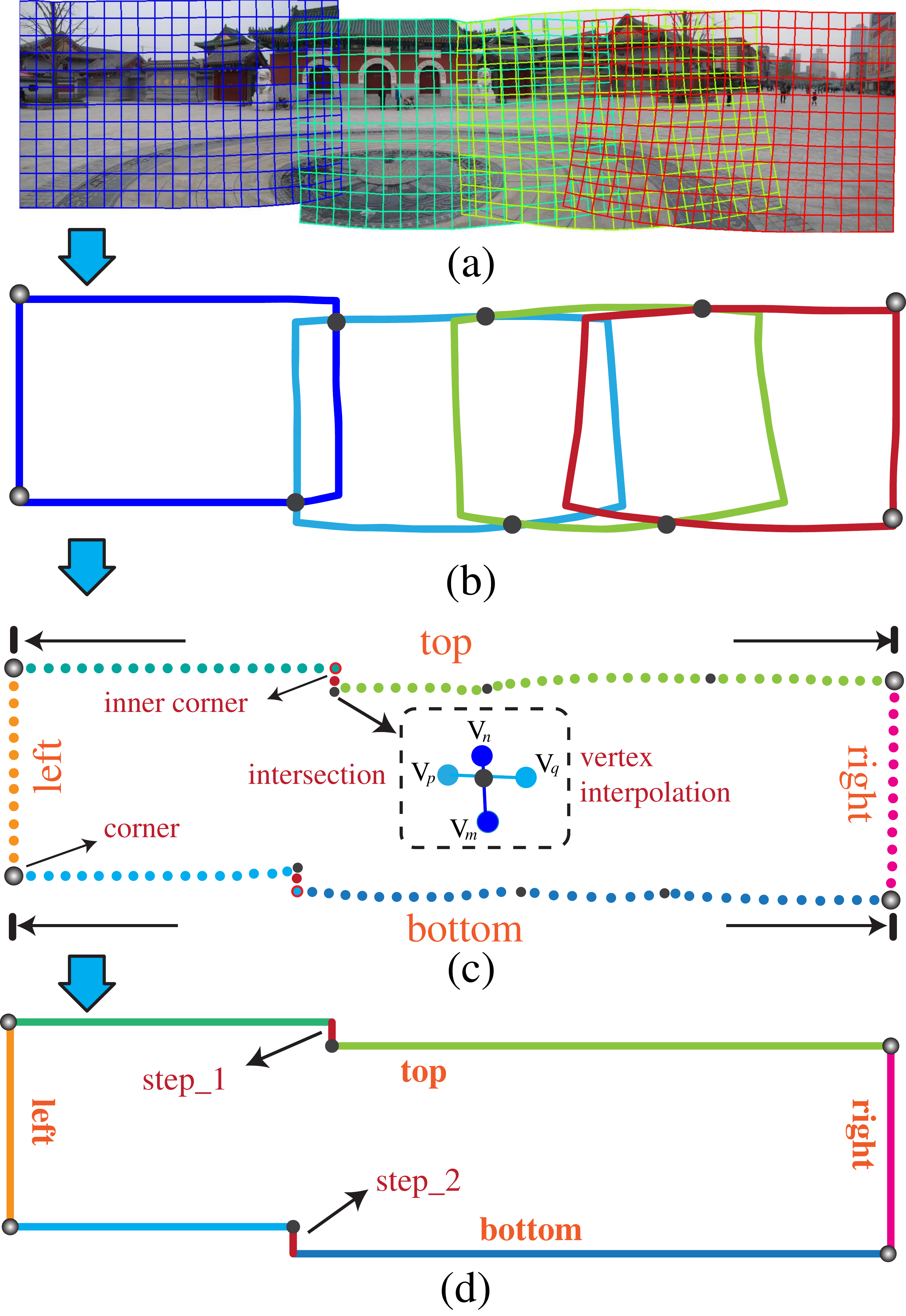}
  \caption{Irregular boundary extraction and piecewise rectangular boundary construction. (a) meshes of initial stitching, (b) boundary Boolean operations, (c) irregular boundary extraction, (d) piecewise rectangular boundary.} \label{fig:irregular}
\end{figure}

 \subsection{Irregular Boundary Extraction}
 \label{sec:irregular_boundary}
The irregular boundary extraction is an important step for panorama rectangling.
Unlike~\cite{journals/tog/HeC013}, which places only one mesh over the stitched panorama with an irregular boundary, our method places separate mesh for each image to be stitched.
As a result, the irregular boundary consists of vertices from different image meshes, around overlapping regions, and edges from different meshes intersect with each other, as shown in Fig.~\ref{fig:irregular}(a).
Although  using a single mesh as in~\cite{journals/tog/HeC013} makes the representation simpler, it has an unavoidable limitation that due to the boundary irregularity, the mesh grid may contain regions out of the stitched images, leading to small holes in the rectangling results, as given in Fig.~\ref{fig:holes_he}(c).
In our method, to cope with multiple meshes and arbitrary overlapping situations, we notice that the overall irregular boundary is formed by boundaries of warped meshes. More specifically, the boundary of each warped mesh is a polygon, and the union of all such polygons forms a compound polygon corresponding to the stitched image. The irregular boundary of the stitching result can be simply obtained
as the boundary of the compound polygon, as illustrated in Fig.~\ref{fig:irregular}(b).

Therefore,  we propose a simple and effective algorithm for irregular boundary extraction, based on the polygon Boolean union operations~\cite{journals/gandc/MartinezRF09}.
The input includes the mesh vertices $\hat{V}^i$ of each warped image $I_i$, and the goal is to obtain the vertices on the irregular boundary. As shown in Fig.~\ref{fig:irregular}(c), to simplify the discussion we assume that the irregular boundary is split into four sides, denoted as $B^k$ ($k=1, 2, 3, 4$) corresponding respectively to \emph{top}, \emph{right}, \emph{bottom} and \emph{left} sides (in the clockwise order). Let $\hat{P}^i$ be the polygon of the $i^{\rm th}$ warped image. We use the algorithm in~\cite{journals/gandc/MartinezRF09} to efficiently calculate the compound polygon $\hat{P}$ as the union of all these image polygons
\begin{equation}
\hat{P} = \bigcup_{i=1}^N \hat{P}^i.
\label{eq:hatP}
\end{equation}
Denote by $\hat{P}_j$ the $j^{\rm th}$ vertex of $\hat{P}$ in the clockwise order. We similarly use it to represent the position of the vertex without ambiguity. $\hat{P}_j$ can either be a boundary vertex from a warped mesh, or an intersection of two warped mesh edges. We introduce an indicator function $\zeta(\hat{P}_j)$, which is $1$ if it is a vertex from a warped mesh, and $0$ otherwise. For the former case, we use $\hat{V}_{k_j}$ to indicate the warped vertex. In the later case, the position of the intersection point is obtained using a linear interpolation of the 4 vertices from the two intersecting grid edges.
Denote by $\boldsymbol{\kappa}_j = [\hat{V}_{m_j},\hat{V}_{n_j},\hat{V}_{p_j},\hat{V}_{q_j}]$ the vector containing 4 vertices, and $\boldsymbol{\eta}_j = [c_{m_j},c_{n_j},c_{p_j},c_{q_j}]$ their contributing weights. The position of the intersection point $\hat{P}_j = \boldsymbol{\kappa}_j \cdot \boldsymbol{\eta}_j$.

To work out the irregular boundary sides $B^k$ ($k=1, 2, 3, 4$), we first obtain the axis-aligned bounding rectangle $\hat{R}$ of $\hat{P}$. Denote by $\hat{C}_k$ the 4 corners of the rectangle $\hat{R}$. The 4 corner vertices $V_{C_k}$ on the warped meshes are then defined as the vertices on the warped meshes closest to $\hat{C}_k$, i.e.
\begin{equation}
V_{C_k} = \argmin_{\hat{V}_j \in \hat{V}} \|\hat{V}_j - \hat{C}_k \|.
\label{eq:Vck}
\end{equation}
The 4 corner vertices $V_{C_k}$ split the compound polygon $\hat{P}$ into 4 sides, which are denoted as $B^k$.

The algorithm is summarized in Alg.~\ref{alg:irregular_boundary}.
As shown in Fig.~\ref{fig:irregular}, the initial image stitching result has an irregular boundary formed by the overlapping of $4$ image meshes.
Fig.~\ref{fig:irregular}(b) shows contours of all meshes, where each contour is shown in a different color, and the black circles are intersections of these contours.
As shown in Fig.~\ref{fig:irregular}(c), after the polygon Boolean union operations, the irregular boundaries are correctly extracted and classified into $4$ sides.

\begin{algorithm}[t]
 \label{alg:irregular_boundary}
     \caption{Irregular boundary extraction}
      \KwIn{Mesh vertices $\hat{V}^i$ of each warped image $I_i$, $ i=1, 2,\ldots, N$;}
      \KwOut{Indexes of boundary vertices $B^k$, $k=1, 2, 3, 4$ corresponding to \emph{top}, \emph{right}, \emph{bottom} and \emph{left} sides of the boundary;}
Let $\hat{P}^i$ be the polygon of $I_i$\;
Calculate $\hat{P}$ using polygon union operators in Equ.~\ref{eq:hatP}\;
\ForEach {$\hat{P}_j \in \hat{P}$}
{
    Use $\zeta(\hat{P}_j)$ to indicate if it is a vertex (1) or an intersection point (0)\;
    \If {$\zeta(\hat{P}_j) == 1$}
    {
        Record the vertex of the warped mesh $\hat{V}_{k_j}$;
    }
    \Else
    {
        Record the relevant vertices and their weights: $\boldsymbol{\kappa}_j = [\hat{V}_{m_j},\hat{V}_{n_j},\hat{V}_{p_j},\hat{V}_{q_j}]$;  $\boldsymbol{\eta}_j = [c_{m_j},c_{n_j},c_{p_j},c_{q_j}]$;
    }
}
Determine the bounding rectangle $\hat{R}$ of $\hat{P}$\;
Find the 4 corners $\hat{C}_k$ of $\hat{R}$\;
Calculate the 4 corner vertices $V_{C_k}$ using Equ.~\ref{eq:Vck}\;
Split $\hat{P}$ into $B^k$ using $V_{C_k}$ ($k=1, 2, 3, 4$);

\end{algorithm}

\subsection{Piecewise Rectangular Boundary Constraints}

Given vertices $B^k_j$ from the irregular boundary side $B^k$ ($k=1, 2, 3, 4$), the aim of this step is to group them to form boundary sections, where each section $S^k_j$ represents a sequence of boundary vertices that are in the same direction and should be aligned horizontally or vertically in the target piecewise rectangular shape, as illustrated in Fig.~\ref{fig:irregular}(d), where each section is shown in a different color. We initialize each adjacent vertex pairs $(B^k_j, B^k_{j+1}) \in  B^k$ as a boundary section $S^k_j$. We then repeatedly merge two adjacent boundary sections $S^k_{j_1}$ and $S^k_{j_2}$ if they are in the same direction, i.e. $dir(S^k_{j_1}) = dir(S^k_{j_2})$, where $dir(\cdot)$ works out the dominant direction as either horizontal (0) or vertical (1). When no further merging is possible under this rule, we further merge very short sections with less than $2$ vertices (referred to as small steps) to their neighboring sections, to avoid overly complicated boundary structure.
After analyzing the irregular boundaries, we calculate the target boundary value of each section $val(S^k_j)$ by averaging their coordinates in the corresponding direction. The algorithm is summarized in Alg.~\ref{alg:piecewise_analysis}.
As shown in Fig.~\ref{fig:irregular}(d), the top and bottom boundary sides contain $3$ segments each, and steps orthogonal to the sides are essential to reduce distortions in panorama rectangling.

\begin{algorithm}[t]
 \label{alg:piecewise_analysis}
     \caption{Piecewise rectangular boundary analysis}
      \KwIn{Irregular boundary sides $B^k$ from Alg.~\ref{alg:irregular_boundary};}
      \KwOut{Boundary sections $S^k = \{ S^k_j\}$ corresponding to the boundary side $B^k$;}
    $S^k = \emptyset$\;
    \ForEach {$B^k_j, B^k_{j+1} \in B^k$}
    {
        Let $S^k_j = \{ B^k_j, B^k_{j+1} \}$\;
        Add $S^k_j$ to $S^k$;
    }
    \Repeat {no further merging}
    {
        \ForEach {adjacent boundary section $(S^k_{j_1}, S^k_{j_2})$ $S^k_{j_1} \bigcap S^k_{j_2} \neq \emptyset$}
        {
            \If {$dir(S^k_{j_1}) == dir(S^k_{j_2})$}
            {
                Merge $S^k_{j_2}$ to $S^k_{j_1}$;
            }
        }

        \ForEach {boundary section $S^k_j$, $|S^k_j| < 2$}
        {
            Merge $S^k_j$ to its previous boundary section;
        }
    }
    \ForEach {boundary section $S^k_j$}
    {
        \If {$dir(S^k_j) == 0$ (horizontal)}
        {
            $val(S^k_j) = Avg(B^k_t.y)  (\forall B^k_t \in S^k_j)$;
        }
        \Else
        {
            $val(S^k_j) = Avg(B^k_t.x)  (\forall B^k_t \in S^k_j)$;
        }
    }

\end{algorithm}

\subsection{Piecewise Rectangular Stitching}\label{sec:opt:piecewise}
We design a global optimization which simultaneously finds the optimal image stitching and piecewise rectangling results.
Our energy function contains terms about feature alignment, shape preserving and global similarity constraints that are used for stitching.
Besides, we also consider regular boundary and straight line preserving constraints that are used for avoiding unexpected distortions when rectangling irregular boundaries.
The energy terms for stitching have been defined in Section~\ref{sec:Initial image stitching}, and we now define energy terms for irregular boundary rectangling as follows.

\textbf{Regular boundary preserving}.
With the piecewise rectangular boundary constraints, we define the regular boundary preserving energy as
\begin{equation} \label{equ:piecewise_boundary}
\begin{split}
   &\phi_r(\hat{V})=\sum\limits_{k=1}^4\sum\limits_{S^k_j \in B^k}\sum\limits_{\hat{V}_t \in S^k_j}\| \boldsymbol{\Lambda}_{dir(S^k_j)}[\zeta(\hat{V}_t) \hat{V}_t+\\
   &(1-\zeta(\hat{V}_t))(\boldsymbol{\kappa}_t \cdot \boldsymbol{\eta}_t)]-val(S^k_j) \|^2,
\end{split}
\end{equation}
where $B^k$, $k=1,2,3,4$ refer to the boundary sides in \emph{top}, \emph{right}, \emph{bottom} and \emph{left} directions;
$S^k_j$ represents a boundary section in the $k^{\rm th}$ side, and $val(\cdot)$ refers to the value of the target boundary section.
As defined before, $\zeta(\hat{V}_t)$ indicates the type of the boundary point, either as a vertex of the meshes (1) or their intersection (0).
$\boldsymbol{\Lambda}_0 = [0 \quad 1]$  
and $\boldsymbol{\Lambda}_1 = [1 \quad 0]$ are
$1 \times 2$ matrices, used to extract the $y$ and $x$ components of the coordinates respectively, to constrain the position of the boundary point to be close to the desired values.


\textbf{Straight line preserving}.
To avoid unexpected distortion after warping, we also need to preserve straight lines in panoramas.
In the initial stitching step, we are only concerned about obtaining the irregular boundary, thus the straight line preserving term is not necessary in that step.
We use the line preserving term from~\cite{journals/cgf/LinLCZ16}, and the line segments are detected using~\cite{journals/pami/GioiJMR10}.
Let $L^i$ be the set of all detected line segments in image $I_i$. For a given line segment $l \in L^i$, assume that it contains $p$ sub-segments, with sample points $l_0, l_1, \dots, l_{p}$.
Each sample point $l_j$ is represented by interpolating vertices of mesh grid that contains $l_j$.
Specifically, $l_j=\mathbf{\hat{V}}^i_{l_j} \cdot \mathbf{\Omega}^i_{l_j}$, where $\mathbf{\hat{V}}^i_{l_j}$ refers to the warped grid vertices, and $\mathbf{\Omega}^i_{l_j}$  are the corresponding weights before warping.
The straight line preserving term is defined such that the position of a sample point $l_j$ should be close to the position obtained by a linear interpolation of two endpoints $l_0$ and $l_{p}$ with weights $(1-j/p)$ and $j/p$. We define the energy term as
\begin{align} \label{equ:line_preserving}
\nonumber   \phi_l(\hat{V}) &= \sum\limits_{i=1}^N\sum\limits_{l \in L_i}\sum\limits_{j=1}^{p-1}\|(1-\frac{j}{p})\hat{\mathbf{V}}^i_{l_0} \cdot  \boldsymbol{\Omega}^i_{l_0} \\
   &+ \frac{j}{p} \hat{\mathbf{V}}^i_{l_{p}} \cdot  \boldsymbol{\Omega}^{i}_{l_{p}} -\hat{\mathbf{V}}^{i}_{l_j} \cdot  \boldsymbol{\Omega}^{i}_{l_j}\|^2.
\end{align}

\textbf{Total energy}.
With the piecewise rectangular boundary and straight line preserving constraints,  
the total energy function for our content-preserving image stitching is defined as
\begin{equation} \label{equ:content_preserving_stitching}
\Phi(\hat{V})=\Phi_{stitch}(\hat{V})+\gamma_r \phi_r(\hat{V})+\gamma_l \phi_l(\hat{V}),
\end{equation}
where $\Phi_{stitch}$ is the stitching energy function defined in Section~\ref{sec:Initial image stitching}, $\gamma_r$ and $\gamma_l$ are weights to control the importance of energy terms. We set $\gamma_r=10^3$ to ensure the regularity of boundaries.
In our experiment, we find that line preserving is more important than local shape preserving, thus $\gamma_l$ is set to a higher $15$ to avoid too much distortion in straight lines.

%

\subsection{Refinement of Piecewise Rectangular Boundary}
\begin{figure*}
  \centering
  \includegraphics[width=0.9\textwidth]{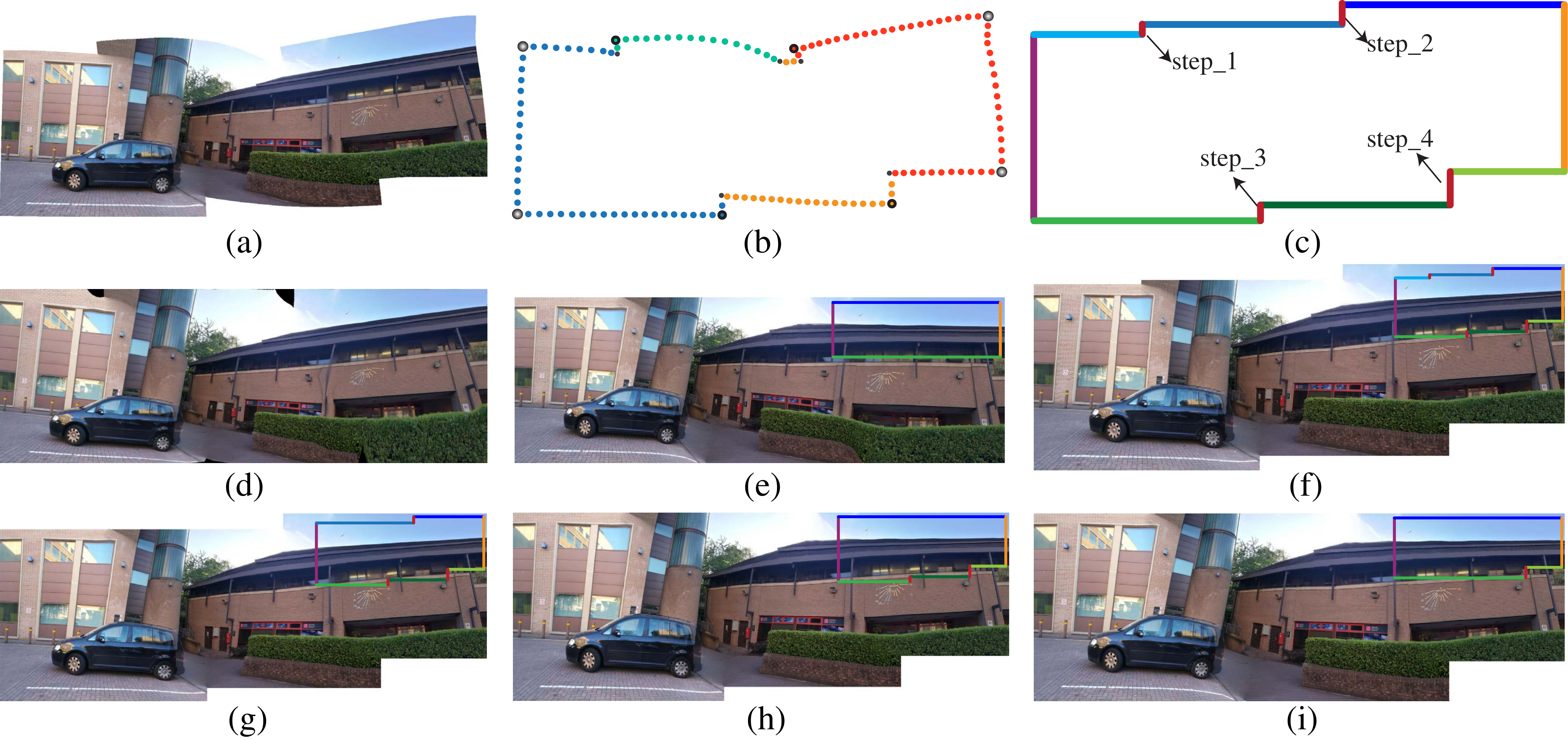}
  \caption{Piecewise rectangling in image stitching. (a) initial stitching result with an irregular boundary, (b) irregular boundary extraction, (c) target boundaries estimation,
  (d) rectangular stitching result by~\cite{journals/tog/HeC013} , (e) our rectangular stitching result, (f-i) stitching results by piecewise rectangling with iterative refinement, and (i) is our final stitching result. } \label{fig:piecewise-process}
\end{figure*}

As shown in Fig.~\ref{fig:piecewise-process}(f), our piecewise rectangular boundary may contain some unnecessary steps, defined as \emph{short} boundary sections orthogonal to the direction of the side,
which may degrade the rectangling effects.
For optimal stitching with regular boundary, we further propose to iteratively refine the piecewise rectangular boundary.
After minimizing the total energy defined in Section~\ref{sec:opt:piecewise}, we calculate the energy $\Phi(\hat{V})$ using the optimized vertices, denoted as $E_0$.
Then, we repeat the following in each iteration until no further improvement can be made.

In the $t^{\rm th}$ iteration (t = 1,2...), we first analyze feature point and line detection results near the boundary sections connected by each step.
If such feature points and lines exist, the corresponding step cannot be removed, see steps in Fig.~\ref{fig:pipeline}(d).
When there are few features and lines, e.g. the local image contains featureless grass and sky, we further analyze such steps as follows: For each step, we attempt to remove it and join its neighboring boundary sections. This leads to a simplified
boundary, and then we apply the same image stitching by minimizing the total energy in Section~\ref{sec:opt:piecewise}. The minimum energy obtained by removing a step is denoted as $E_t$.
We further compare $E_t$ with $E_{t-1}$ from the last iteration.
If $E_t-E_{t-1} < \sigma$, which means that the distortion in this iteration is acceptable, and we accept the new result and proceed to the next iteration. Otherwise, the new result is rejected, and we return the result from the last iteration as the final result.
In this paper, we set the threshold $\sigma$ to $|E_t-E_{t-1}|/20$, which works well in most examples.
Our method is general in that the panorama rectangling proposed by He et al.~\cite{journals/tog/HeC013} can be classified as a special case of our piecewise rectangling, when there are no steps in the target boundary.

Fig.~\ref{fig:piecewise-process}(e) is the rectangling result by our method when all steps are removed, and there exists too much distortion in the bottom-right corner.
Compared with the result by~\cite{journals/tog/HeC013} in Fig.~\ref{fig:piecewise-process}(d) which contains large holes and distortions, our rectangling result is more reasonable.
Figs.~\ref{fig:piecewise-process}(f-i) show results of our piecewise rectangling in each iteration, and the top-right corner of each result shows the shape of the target regular boundary.
These results demonstrate that each iteration makes the boundary of panorama closer to a rectangle, and finally we get the panorama with optimal piecewise rectangular boundary without noticeable distortions, see Fig.~\ref{fig:piecewise-process}(i).

\subsection{Optimization and Result Generation}\label{sec:opt:gen}
For initial image stitching, we first minimize $\Phi_{stitch}(\hat{V})$ defined in Equ.~\ref{equ:global_stitching}, which is global translation invariant.
To ensure a unique solution, we fix the first vertex of the first mesh.
Note that each energy term is quadratic and variables are mesh vertices of each image, and therefore the energy function can be efficiently minimized by solving a sparse linear system.
Since this stitching step is only used to get the target rectangle and irregular boundary, we do not need to render the stitching result by warping and blending.


After irregular boundary extraction, we minimize the total energy defined in Equ.~\ref{equ:content_preserving_stitching} which incorporates the regular boundary and straight line constraints into the stitching framework, thus can simultaneously optimize both stitching and boundary regularity. 
Since both the added terms are quadratic, Equ.~\ref{equ:content_preserving_stitching} can also be efficiently minimized.

With the optimized vertices of each mesh, we further warp each image by texture mapping and suppress seams between different meshes by multibanded blending~\cite{journals/tip/ZhuLWZMLH18}.
For efficiency, we can also simply apply linear blending, which works well in most cases.
Fig.~\ref{fig:pipeline}(f) is the stitching result constrained by our piecewise rectangular boundary. Compared with traditional stitching in Fig.~\ref{fig:pipeline}(b) and existing rectangling method~\cite{journals/tog/HeC013} in Fig.~\ref{fig:pipeline}(g), our method makes a better balance between the distortion and boundary regularity, and preserve panorama contents in a rectangular window as much as possible.

 \begin{figure*}[t] 
  \centering
  \includegraphics[width=0.9\textwidth]{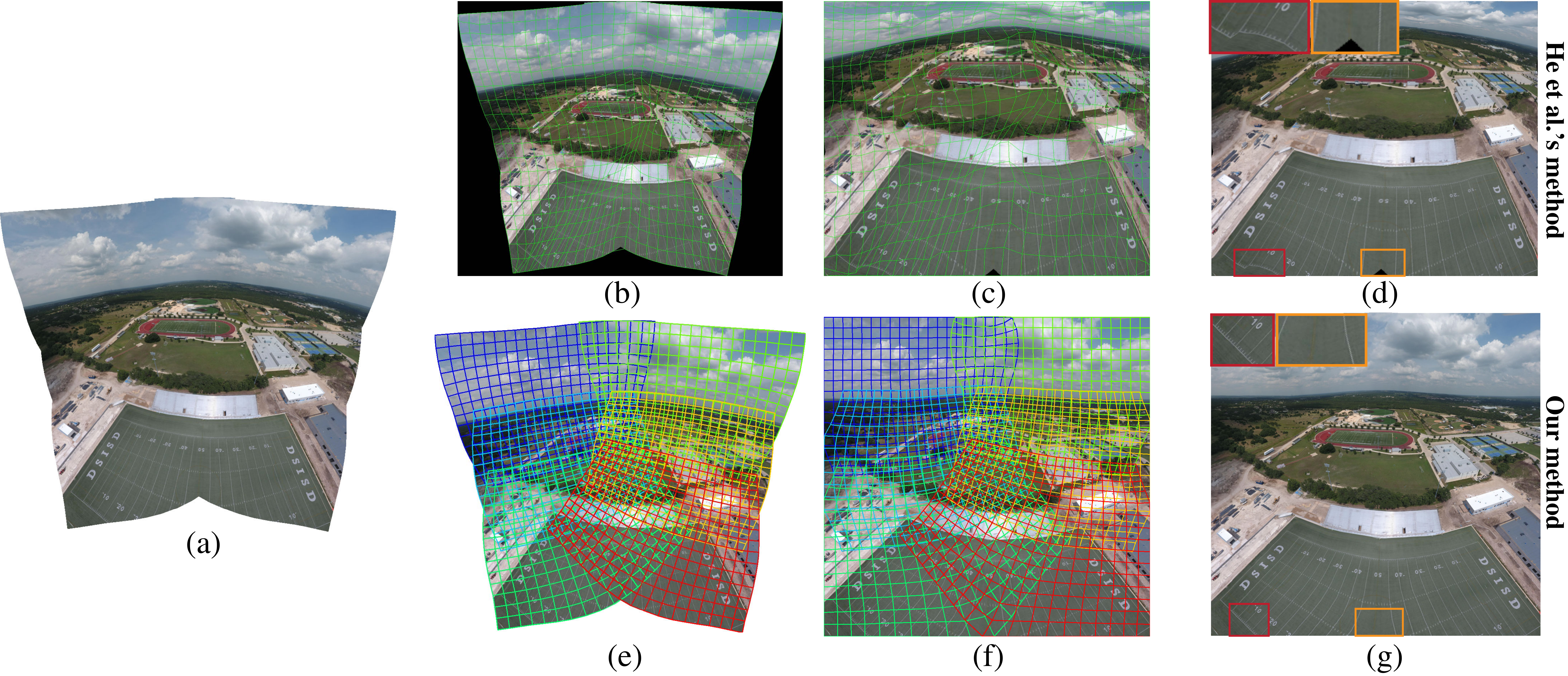}
  \caption{Comparison with He et al.'s~\cite{journals/tog/HeC013} method. (a) the initial stitching result, which is also used as the input to the method in~\cite{journals/tog/HeC013}. Results by~\cite{journals/tog/HeC013}: (b) mesh of the initial stitching, (c) mesh after global warping, (d) rectangular panorama by~\cite{journals/tog/HeC013}.
  Results by our method: (e) meshes of initial stitching, (f) meshes after the global warping, (g) our rectangular panorama.} \label{fig:holes_he}
\end{figure*}

 \begin{figure*}[t] 
  \centering
  \includegraphics[width=.9\textwidth]{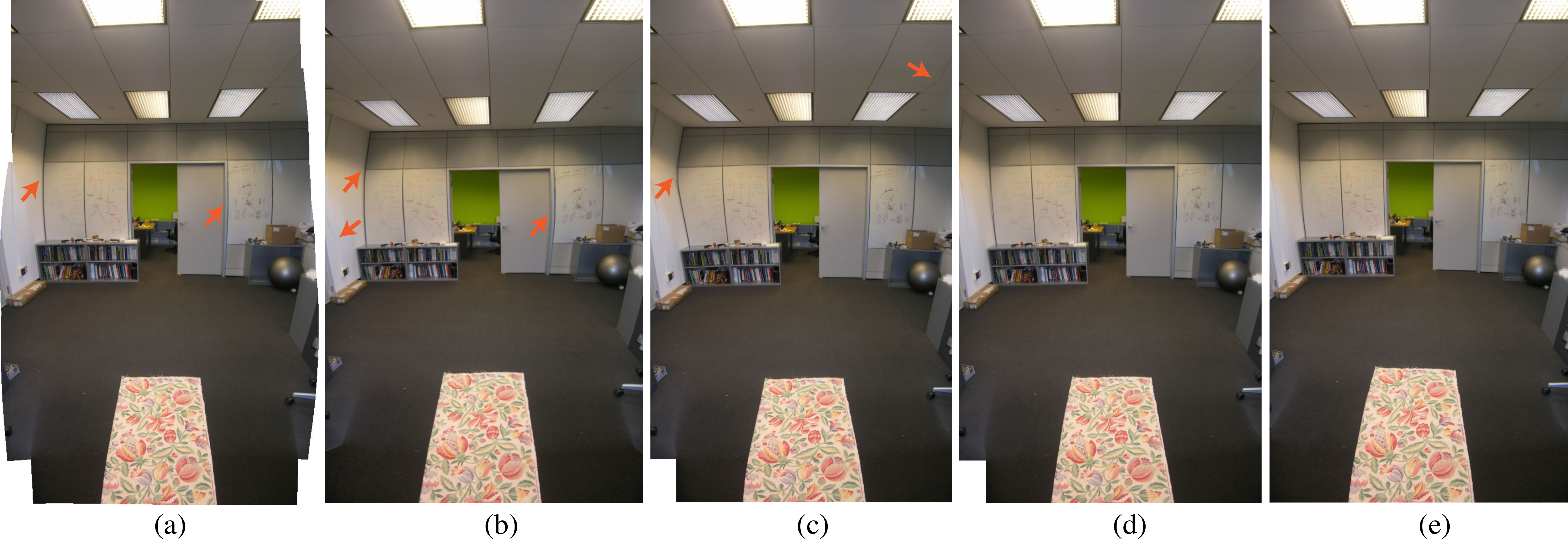}
  \caption{Comparison with state-of-the-art methods. (a) stitching result by~\cite{conf/eccv/ChenC16}, (b) rectangling stitching by~\cite{journals/tog/HeC013}, (c) and (d) are our piecewise rectangling results in the $1^{\rm st}$ iteration with and without line preserving), (e) our final stitching results after several iterations. } \label{fig:result_office}
\end{figure*}

\begin{figure*}[t]
  \centering
  \includegraphics[width=0.75\textwidth]{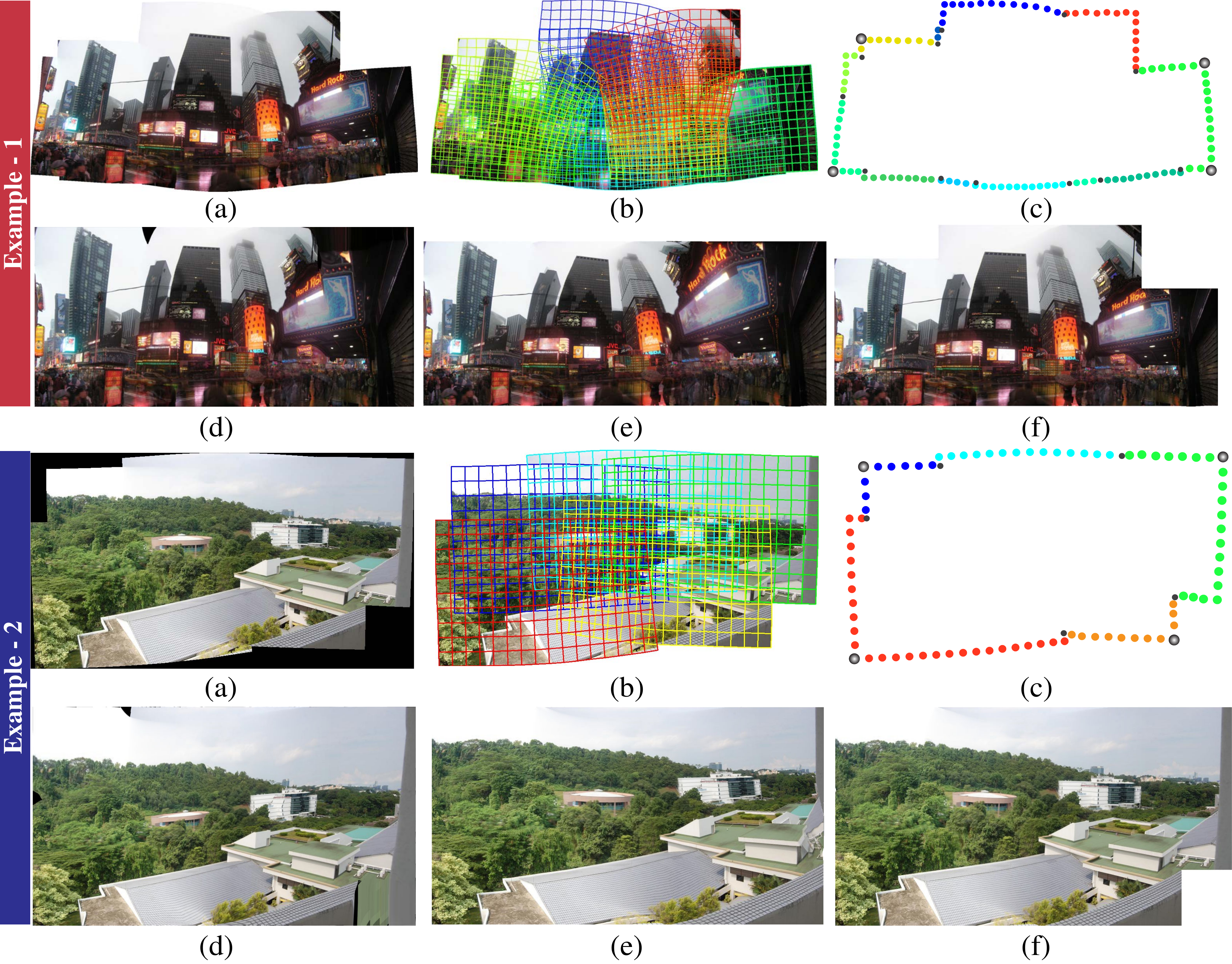}
  \caption{Results and comparisons of stitching with large missing contents. Two examples are presented as follows: (a) initial stitching result with an irregular boundary, (b) warped meshes of initial stitching, (c) irregular boundary extraction, (d) and (e) rectangling stitching by~\cite{journals/tog/HeC013} and our method respectively, (f) our piecewise rectangular stitching result.} \label{fig:piecewise-compare}
\end{figure*}

 \section{RESULTS AND APPLICATIONS}

In this section, we show a variety of panoramic images generated by our image stitching with regular boundary constraints, and provide qualitative comparisons with state-of-the-art methods.
Unlike previous image stitching methods~\cite{journals/pami/ZaragozaCTBS14,journals/tmm/LiWLZZ18}, which focus on more accurate alignment for better stitching, we aim to produce stitched panoramas with regular boundary, thus, similar to \cite{journals/tog/HeC013}, standard quantitative evaluations are not meaningful.
Then, we further give several applications that benefit from our proposed rectangling.
Finally, we report performance and discuss limitations of our method.
In this paper, we use the datasets provided by Chen et al.~\cite{conf/eccv/ChenC16} for image stitching, and Perazzi et al.~\cite{journals/cgf/PerazziSZKWWG15} for video stitching.
For clearer presentation, we only provide input for examples shot by ourselves.

 \subsection{Results and Comparisons}
Fig.~\ref{fig:holes_he} shows a comparison of our method with~\cite{journals/tog/HeC013} for producing rectangular panoramas.
Fig.~\ref{fig:holes_he}(a) is the initial stitching result in the first step of our method.
For fair comparison, we also take it as the input to He et al.'s method~\cite{journals/tog/HeC013}.
In~\cite{journals/tog/HeC013}, a single mesh is placed on the initial stitched panorama with an irregular boundary, 
and it is common that the mesh contains regions out of the stitched panorama,  see Fig.~\ref{fig:holes_he}(b).
As given in Figs.~\ref{fig:holes_he}(c)(d), after global warping, rectangling result by~\cite{journals/tog/HeC013}  may contain \emph{holes}, which degrades the quality of the final rectangular panorama. In addition, method in~\cite{journals/tog/HeC013}  treats stitching and rectangling as two individual processes, thus cannot well preserve the local and global structures of the scene.
Compared with~\cite{journals/tog/HeC013} , we utilize the meshes (see Fig.~\ref{fig:holes_he}(e)) from the initial stitching step, thus can avoid the \emph{hole} problem entirely.
With these meshes, a global optimization which combines stitching and rectangling constraints is constructed, and the final result can not only obtain a regular boundary, but also well preserve local and global structures, see Figs.~\ref{fig:holes_he}(f)(g). 

Fig.~\ref{fig:result_office} gives comparison with state-of-the-art methods in terms of line preserving.
In~\cite{conf/eccv/ChenC16}, the line segment detection is used for feature preserving in global transformation, like scale and rotation, thus cannot well preserve straight lines, as shown in (a).
Using the stitched panorama from (a), method in~\cite{journals/tog/HeC013} is limited by the input, thus fails to preserve straight lines.
(c) and (d) are our initial piecewise rectangular stitching results without and with line preserving, and the results show that our method can well preserve straight lines. Arrows in (a), (b), (c) point out regions that fail to preserve straight lines.
(e) is our final stitching result after several iterations, which not only preserves lines, but also provides a panorama with a rectangular boundary.

Fig.~\ref{fig:piecewise-compare} presents results and comparisons of stitching for scenes with large missing contents.
We provide two examples to show the effectiveness of our method for such challenging cases.
For each example, (a) gives the initial stitching result,  which is also used as the input to He et al.'s method~\cite{journals/tog/HeC013}.
 (b) and (c) show the meshes after initial stitching and the extracted irregular boundaries respectively.
(d) and (e) are the rectangular stitching results by~\cite{journals/tog/HeC013} and our method.
Although both of them have distortions, our result is more reasonable and visually pleasing.
In addition,  due to the drawback of the mesh representation in~\cite{journals/tog/HeC013}, the warped panoramas contain holes.
With the optimized piecewise rectangular boundary, our result in (f) has unnoticeable distortions, while preserving its content in a rectangular windows as much as possible.

\begin{figure}[t] 
  \centering
  \includegraphics[width=0.5\textwidth]{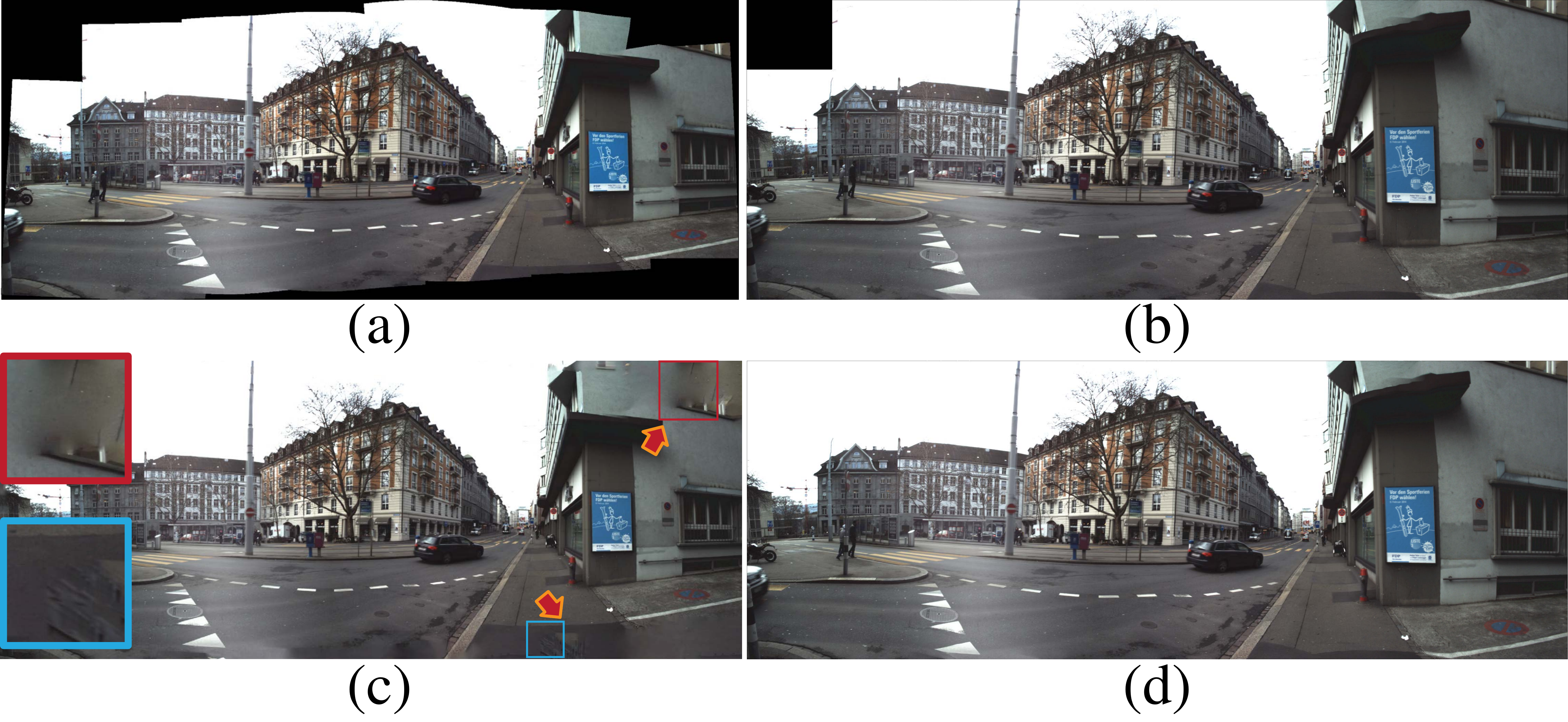}
  \caption{Comparison with image completion. (a) initial stitching result by~\cite{conf/eccv/ChenC16}, (b) our piecewise rectangling panorama, (c) image completion result by~\cite{journals/tog/HuangKAK14} applied to (a), (d) image completion result by applying~\cite{journals/tog/HuangKAK14} to our stitching result in (b). } \label{fig:completion}
\end{figure}

An alternative approach to generating rectangular panoramas is image completion. 
Fig.~\ref{fig:completion} compares panorama completion results, using traditional stitching and our method as input.
(a) is the stitching result by~\cite{conf/eccv/ChenC16}, which has irregular boundary and large content missing.
By completing holes in (a) using the method in~\cite{journals/tog/HuangKAK14}, we get the rectangular panorama shown in (c). The close-up windows show that the completion result is poor in synthesizing semantic content.
(b) is the piecewise rectangling result by our method, which preserves regular boundaries while preventing undesirable distortions.
Based on our result, it is much easier for image completion to synthesize regular holes in the top left corner. 
The result in (d) shows that the combination of our method and image completion is successful.

\begin{figure}[t] 
  \centering
  \includegraphics[width=0.5\textwidth]{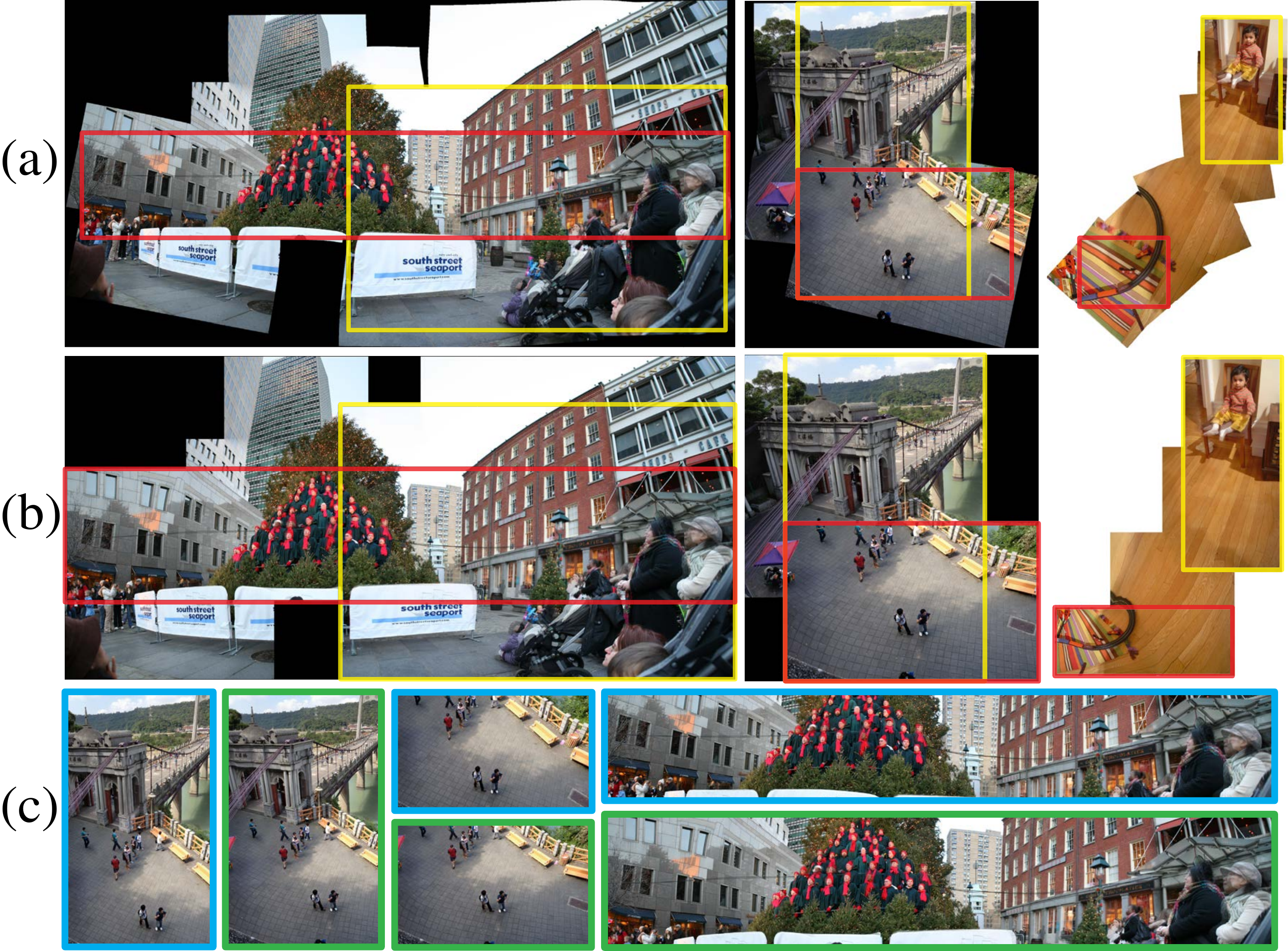}
  \caption{Results of challenging cases. (a) Initial stitching results with irregular boundaries, (b) results of our piecewise rectangling stitching, (c) cropped images based on the stitched panorama by~\cite{conf/eccv/ChenC16}(cyan border) and our results (green border).} \label{fig:challenging_cases}
\end{figure}

Fig.~\ref{fig:challenging_cases} gives results of challenging cases, which contain a large amount of missing content, thus previous panorama rectangling method~\cite{journals/tog/HeC013} cannot produce plausible results.
Line (a) shows stitching results by~\cite{conf/eccv/ChenC16}  with irregular boundaries, and Line (b) shows our piecewise rectangling panoramas.
\emph{Red} and \emph{yellow} rectangles in Line (b) show possible cropping results from the results by~\cite{conf/eccv/ChenC16} and our method.
Line (c) further gives the final cropped panoramas by~\cite{conf/eccv/ChenC16}  and our method.
It is obvious that, with the generated piecewise rectangular boundaries using our method, the panoramic images can be easily cropped or even completed,
and we can obtain panoramic images with more content and without noticeable distortions by choosing a rectangular window in our results.
Thus, compared with traditional stitching with irregular boundaries, our method is effective to improve the visual effects and viewing experience of panoramic images.

\begin{figure*} 
  \centering
  \includegraphics[width=0.75\textwidth]{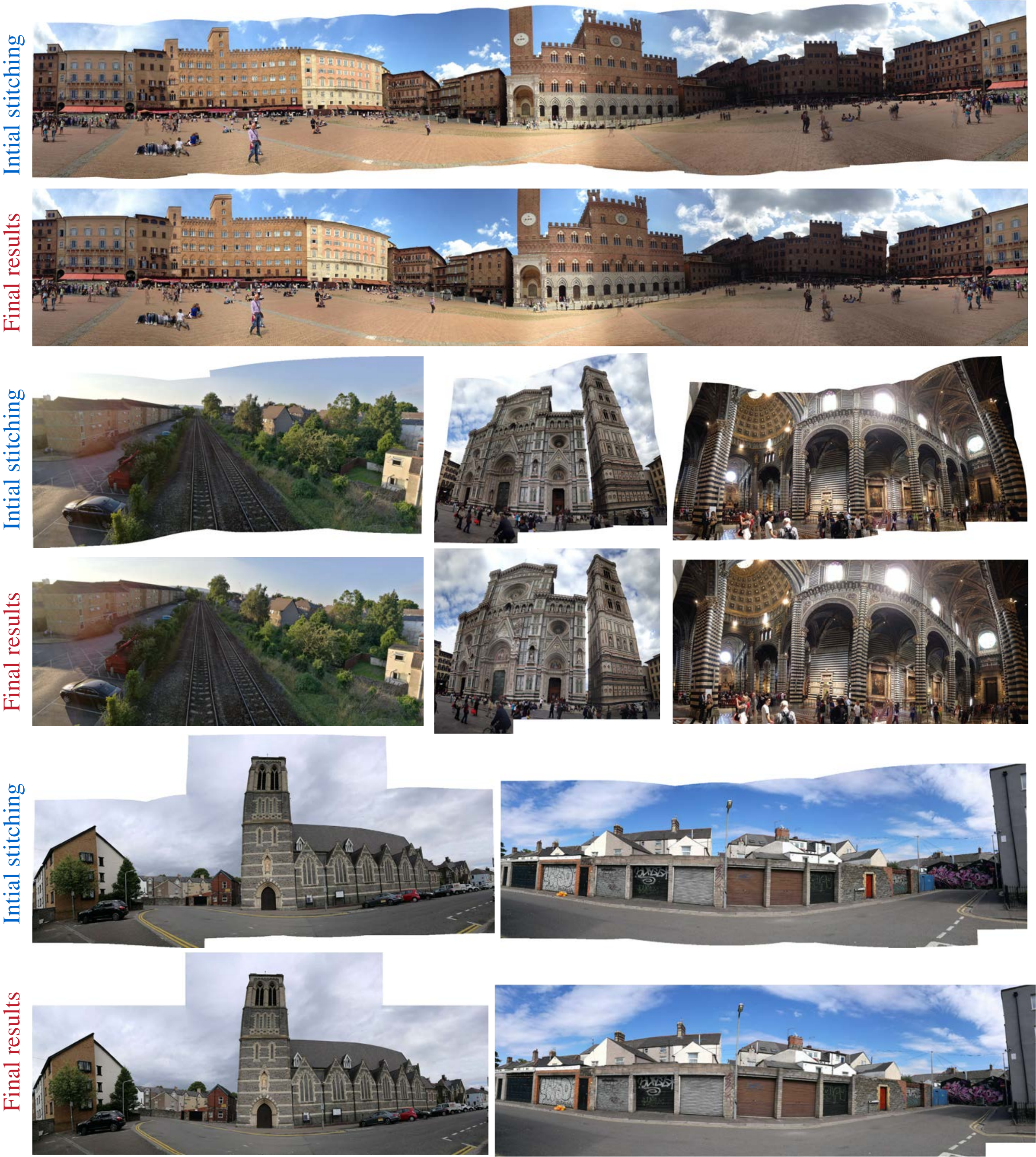}
  \caption{More results. The \emph{initial stitching} are results without regular boundary constraints, and \emph{final results} are obtained by our method.} \label{fig:more_results}
\end{figure*}

Fig.~\ref{fig:more_results} gives more results using our method. Compared with results by \emph{initial stitching}, the \emph{final results} by our method can produce panoramas with regular boundaries, which can provide better wide viewing experience, and preserve more image content in a rectangular window.

 \subsection{Applications}

 \subsubsection{Selfie expansion}

Selfies have become increasingly more popular in recent years with the fast development of smart phones and tablets. They are typically shot by holding the mobile device in a hand or with a selfie stick. Since the camera is very close to the person, selfie photos always have limited field-of-view, which reduces its fun.
We demonstrate that our method can be used to produce selfie panoramas with large field-of-view.
The front camera designed for selfie shooting cannot be used for shooting a panorama.
Thus we first take photos of the panorama view using the back camera, and then shoot the selfie portrait using the front camera facing the background of the panorama.
Our method stitches these images to form a selfie with large field-of-view.

\begin{figure} 
  \centering
  \includegraphics[width=0.48\textwidth]{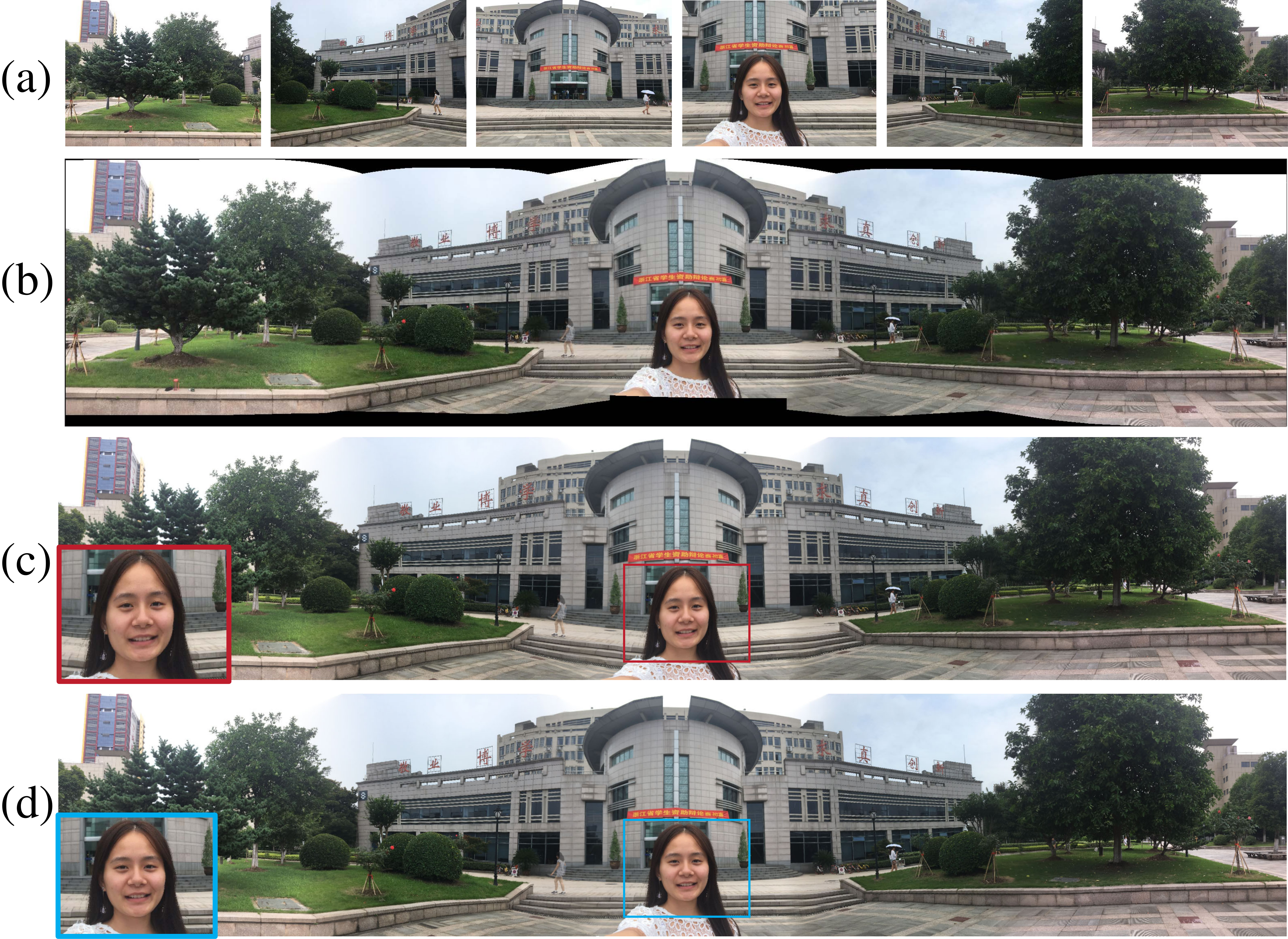}
  \caption{Application to selfie expansion. (a) input photos, (b) initial stitched image with an irregular boundary, (c) result of our method with a regular boundary, which distorts the human face, (d) result of our method with a regular boundary and face preservation, which can avoid the unwanted face distortion. } \label{fig:selfie}
\end{figure}

We further propose to preserve the important visual features on the face region during stitching to avoid distortion on faces.
We first detect the face from the portrait photo, and modify Equ.~\ref{equ:shape_preserving} (shape consistency term) as
\begin{equation} \label{equ:shape_preserving1}
\begin{split}
\phi_s(\hat{V}) = \sum\limits_{i=1}^N\sum\limits_{\hat{V}^i_j \in \hat{V}^i} \
    \alpha_j^i \|\hat{V}_{j}^i -\hat{V}_{j_1}^i-\xi \mathbf{R}(\hat{V}_{j_0}^i -\hat{V}_{j_1}^i)||^2\
\end{split}
\end{equation}
where $\alpha_j^i$ refers to the saliency value of vertex $\hat{V}_j^i $. A larger value ($\alpha_j^i=20$) is specified for vertices in the face region, and $1$ otherwise.
By preserving the shape of the mesh in the face region, the stitching results are more visually pleasing.
An example is shown In Fig.~\ref{fig:selfie}. (a) shows all the input images, including photos for the panorama background and the portrait photo.
We can see that result by~\cite{conf/eccv/ChenC16} in (b) contains irregular boundaries.
(c) is the result by our piecewise rectangular stitching method without considering face features, where the portrait is distorted too much.
As shown in (d), the proposed method with constraints on preserving facial regions  generates a better selfie panorama.

\subsubsection{Rectangling video panoramas}
\begin{figure*} 
  \centering
  \includegraphics[width=0.85\textwidth]{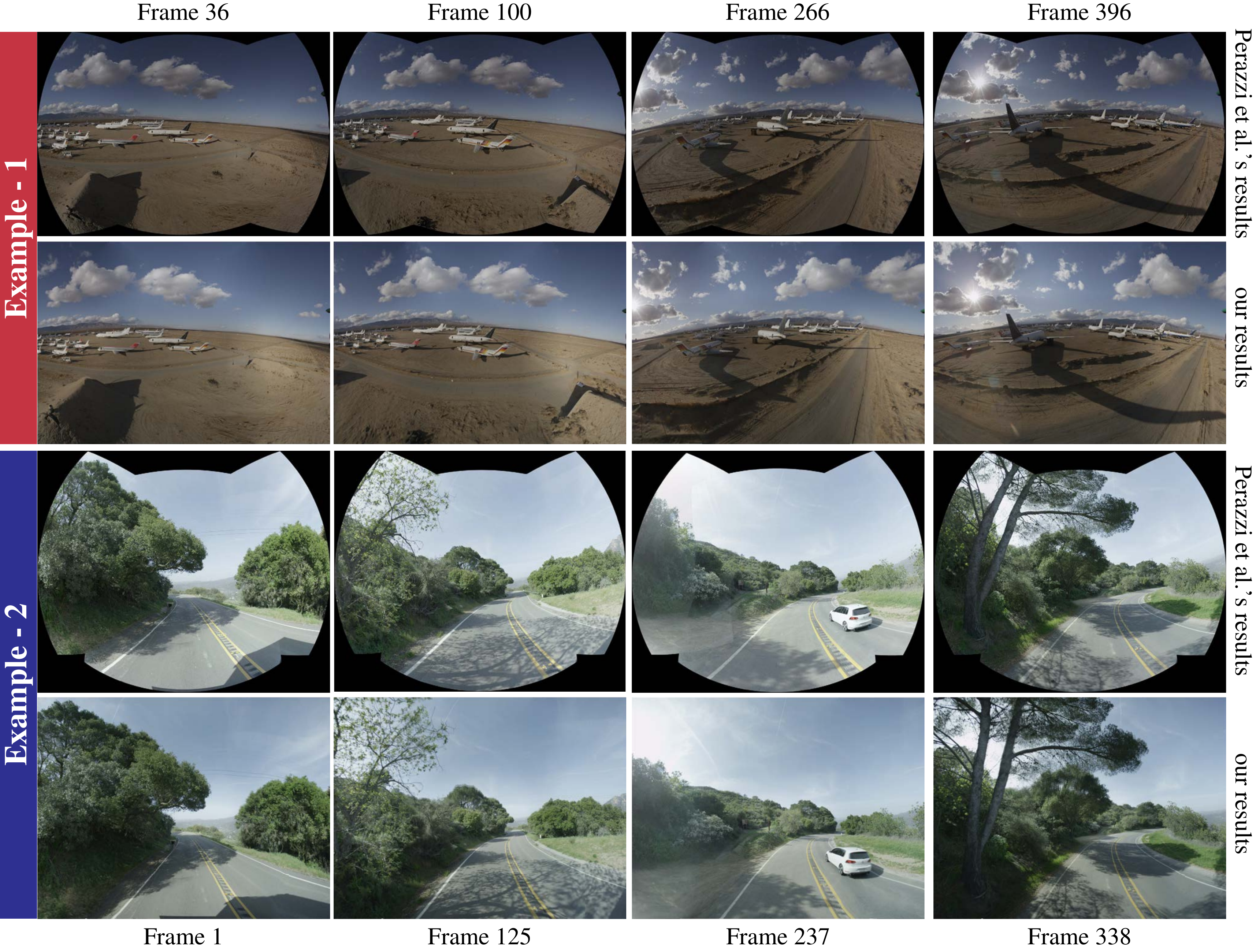}
  \caption{Application to rectangling video panoramas. We give two examples, and each example shows stitched panorama of $4$ different frames by~\cite{journals/cgf/PerazziSZKWWG15} and our method respectively.} \label{fig:video}
\end{figure*}

We further apply our method to video stitching.
In fact, it is difficult to stitch videos from individual hand-held cameras, and rectangling them is even more challenging.
The reason is that the regular boundary in each frame would be different and the temporal coherence is difficult to maintain due to the shaking in each video.
Inspired by~\cite{journals/cgf/PerazziSZKWWG15}, we aim to produce rectangular panoramic videos from unstructured camera arrays with \emph{fixed} camera configurations. This is more manageable, as the warping parameters for stitching individual frames are nearly invariant.
For temporal coherence, we propose a simple and effective scheme as follows:
We first divide a video into several blocks ($35$ frames per block in our experiments with neighboring blocks having overlaps of $15$ frames).
For each block, we compute the stitched panorama for the first frame, and the warping parameters are used for the other frames in the block.
For the overlapping part, the warping parameters are a linear combination of neighboring blocks, gradually transitioning from the first set of parameters to the second.
Fig.~\ref{fig:video} shows two sets of results, and each set shows video panoramas of different frames by~\cite{journals/cgf/PerazziSZKWWG15} and our method.
The comparison shows that our method is effective in rectangling video panoramas shot by fixed camera arrays.
Please refer to the supplementary video for results of our panoramic videos, and comparison with~\cite{journals/cgf/PerazziSZKWWG15}.

 \subsection{Performance}

 We report performance of our method on an Intel Core i7 8550U 1.8GHz laptop with 16G RAM.
 Take Fig.~\ref{fig:pipeline} as an example, the input contains $5$ images ($800\times600$), and the total time is about $3.5$ second.
  The initial stitching costs $0.81$ second, which includes feature matching, line detection, energy construction and optimization.
Then, the stitching with rectangular boundary constraints costs $0.49$ second, which includes the irregular boundary extraction, boundary constraint construction and iterative optimization.
Finally, with the warped vertices, texture mapping and blending are performed, and the time cost is $2.17$ second.
In the optimization, most energy terms are similar, thus we construct them only once.
In addition, since all energy terms are quadratic, the optimization can be efficiently solved.
For our iterative optimization in the piecewise rectangling, results in each iteration are similar, thus we takes the result of last iteration as initialization, and apply the conjugate gradient method to make the optimization more efficiently.

For high resolution images, we first downsample each image to a fixed size ($0.5$ Mega-pixel), and the initial stitching and warping are performed on these downsampled images.
Then we upsample the warped vertices through bilinear interpolation, and the final results are obtained by efficient texture mapping and blending on the original high resolution images.

  \subsection{Limitations}
\begin{figure} 
  \centering
  \includegraphics[width=0.5\textwidth]{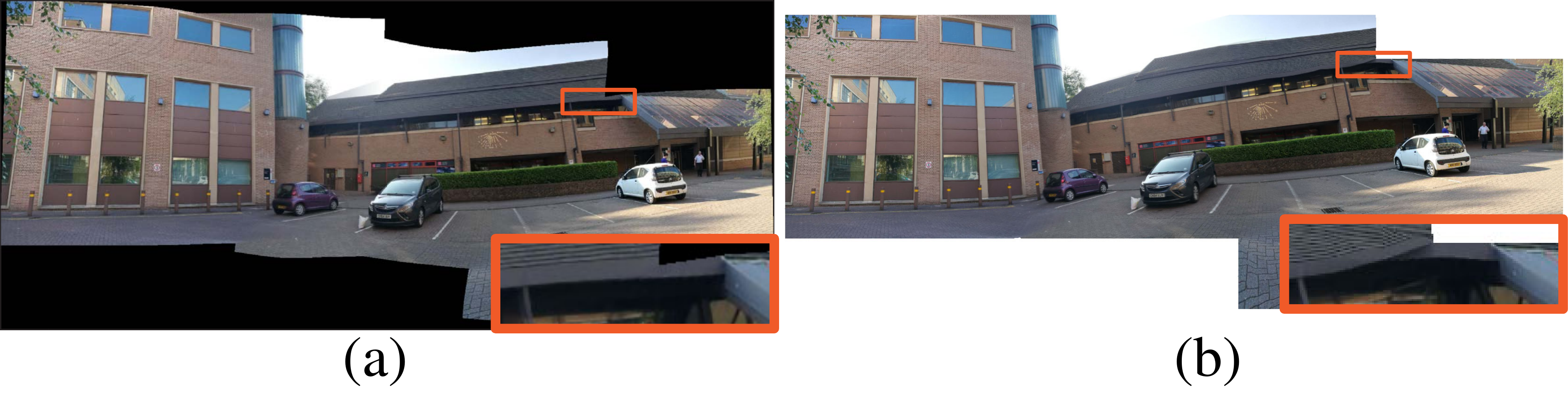}
  \caption{Failure case: when there is strong structure in the intersection of meshes, our method may fail to preserve the structure.} \label{fig:failure}
\end{figure}

Due to the free movement of hand-held cameras, panoramic images inevitably have irregular boundaries and missing content.
Our piecewise rectangling stitching can effectively rectify these problems by warping-based optimizations with regular boundary constraints.
However, there are still some limitations:
(1) Similar to most warping-based methods, our method cannot preserve all lines well when there are many lines in local regions.
(2) Our method may fail when there is strong structure near the intersection of neighboring meshes.
See Fig.~\ref{fig:failure} for an example, where the zoom-in view shows that, our piecewise rectangling scheme may introduce unwanted distortion in order to preserve the piecewise rectangular boundaries.

 \section{CONCLUSION}

In this paper, we have proposed an efficient approach for content-preserving stitching with regular boundary constraints, which can generate panoramic images with regular boundaries.
Our main contribution is a global optimization which incorporates the regular boundary constraints in the framework of image stitching.
Based on the traditional stitching with irregular boundaries, we analyze the warped meshes and extract the outer irregular boundary, 
and then setup the piecewise rectangular boundary constraints for the optimization to get the final content-preserving stitching result.
Experimental results and comparisons show that our method is effective and outperforms state-of-the-art methods.
Especially for panoramic scenes with missing contents, our piecewise rectangling can not only regularize the stitching boundary as much as possible, but also avoid unwanted distortions.
Some challenging examples show the robustness and practicability of our method.
We further apply our method to selfie expansion and video stitching, which demonstrate the versatility of our approach.

In the future, we will consider more features to improve the performance of panorama rectangling, such as visual saliency, scene content etc.
For video stabilization and stitching, the warping-based method may also introduce irregular boundaries.
Regularizing the boundary of warped videos can preserve more content in a cropping window and improve the viewing experiences.
However, for videos shot by freely moving hand-held cameras, it is difficult to define the regular boundary constraints, and maintain the spatial-temporal coherence.
We leave these problems as our future work.
%
%
%
%


%

\ifCLASSOPTIONcaptionsoff
  \newpage
\fi


\begin{thebibliography}{1}
\bibitem{journals/tog/HeC013}
K.~He, H.~Chang, and J.~Sun, ``Rectangling panoramic images via warping,''
  \emph{{ACM} Trans. Graph.}, vol.~32, no.~4, pp. 79:1--79:10, 2013.

\bibitem{journals/mta/YenYC17}
S.~Yen, H.~Yeh, and H.~Chang, ``Progressive completion of a panoramic image,''
  \emph{Multimedia Tools Appl.}, vol.~76, no.~9, pp. 11\,603--11\,620, 2017.

\bibitem{journals/tog/BarnesSFG09}
C.~Barnes, E.~Shechtman, A.~Finkelstein, and D.~B. Goldman, ``Patchmatch: a
  randomized correspondence algorithm for structural image editing,''
  \emph{{ACM} Trans. Graph.}, vol.~28, no.~3, pp. 24:1--24:11, 2009.

\bibitem{conf/eccv/ChenC16}
Y.~Chen and Y.~Chuang, ``Natural image stitching with the global similarity
  prior,'' in \emph{Computer Vision - {ECCV} 2016 - 14th European Conference,
  Amsterdam, The Netherlands, October 11-14, 2016, Proceedings, Part {V}},
  2016, pp. 186--201.

\bibitem{journals/ftcgv/Szeliski06}
R.~Szeliski, ``Image alignment and stitching: {A} tutorial,'' \emph{Foundations
  and Trends in Computer Graphics and Vision}, vol.~2, no.~1, 2006.

\bibitem{journals/ijcv/BrownL07}
M.~Brown and D.~G. Lowe, ``Automatic panoramic image stitching using invariant
  features,'' \emph{International Journal of Computer Vision}, vol.~74, no.~1,
  pp. 59--73, 2007.

\bibitem{conf/CVPR/GaoKB11}
J.~Gao, S.~J. Kim, and M.~S. Brown, ``Constructing image panoramas using
  dual-homography warping,'' in \emph{IEEE Conference on Computer Vision and
  Pattern Recognition (CVPR)}, 2011, pp. 49--56.

\bibitem{journals/pami/ZaragozaCTBS14}
J.~Zaragoza, T.~Chin, Q.~Tran, M.~S. Brown, and D.~Suter,
  ``As-projective-as-possible image stitching with moving {DLT},'' \emph{{IEEE}
  Trans. Pattern Anal. Mach. Intell.}, vol.~36, no.~7, pp. 1285--1298, 2014.

\bibitem{conf/CVPR/LinPRA15}
C.~C. Lin, S.~U. Pankanti, K.~N. Ramamurthy, and A.~Y. Aravkin, ``Adaptive
  as-natural-as-possible image stitching,'' in \emph{IEEE Conference on
  Computer Vision and Pattern Recognition (CVPR)}, 2015, pp. 1155--1163.

\bibitem{conf/ICCV/LiY0Q15}
S.~Li, L.~Yuan, J.~Sun, and L.~Quan, ``Dual-feature warping-based motion model
  estimation,'' in \emph{2015 {IEEE} International Conference on Computer
  Vision, {ICCV} 2015, Santiago, Chile, December 7-13, 2015}, 2015, pp.
  4283--4291.

\bibitem{conf/cvpr/ChangSC14}
C.~Chang, Y.~Sato, and Y.~Chuang, ``Shape-preserving half-projective warps for
  image stitching,'' in \emph{2014 {IEEE} Conference on Computer Vision and
  Pattern Recognition, {CVPR} 2014, Columbus, OH, USA, June 23-28, 2014}, 2014,
  pp. 3254--3261.

\bibitem{conf/cvpr/ZhangL14a}
F.~Zhang and F.~Liu, ``Parallax-tolerant image stitching,'' in \emph{2014
  {IEEE} Conference on Computer Vision and Pattern Recognition, {CVPR} 2014,
  Columbus, OH, USA, June 23-28, 2014}, 2014, pp. 3262--3269.

\bibitem{conf/eccv/LinJCDL16}
K.~Lin, N.~Jiang, L.~Cheong, M.~N. Do, and J.~Lu, ``{SEAGULL:} seam-guided
  local alignment for parallax-tolerant image stitching,'' in \emph{Computer
  Vision - {ECCV} 2016 - 14th European Conference, Amsterdam, The Netherlands,
  October 11-14, 2016, Proceedings, Part {III}}, 2016, pp. 370--385.

\bibitem{journals/tmm/LiWLZZ18}
J.~Li, Z.~Wang, S.~Lai, Y.~Zhai, and M.~Zhang, ``Parallax-tolerant image
  stitching based on robust elastic warping,'' \emph{{IEEE} Trans. Multimedia},
  vol.~20, no.~7, pp. 1672--1687, 2018.

\bibitem{journals/sensors/HeY16}
B.~He and S.~Yu, ``Parallax-robust surveillance video stitching,''
  \emph{Sensors}, vol.~16, no.~1, p.~7, 2016.

\bibitem{journals/itiis/YinLWLZ14}
X.~Yin, W.~Li, B.~Wang, Y.~Liu, and M.~Zhang, ``A novel video stitching method
  for multi-camera surveillance systems,'' \emph{{TIIS}}, vol.~8, no.~10, pp.
  3538--3556, 2014.

\bibitem{journals/cgf/PerazziSZKWWG15}
F.~Perazzi, A.~Sorkine{-}Hornung, H.~Zimmer, P.~Kaufmann, O.~Wang, S.~Watson,
  and M.~H. Gross, ``Panoramic video from unstructured camera arrays,''
  \emph{Comput. Graph. Forum}, vol.~34, no.~2, pp. 57--68, 2015.

\bibitem{journals/computer/AnguelovDFFLLOVW10}
D.~Anguelov, C.~Dulong, D.~Filip, C.~Fr{\"{u}}h, S.~Lafon, R.~Lyon, A.~S.
  Ogale, L.~Vincent, and J.~Weaver, ``Google street view: Capturing the world
  at street level,'' \emph{{IEEE} Computer}, vol.~43, no.~6, pp. 32--38, 2010.

\bibitem{journals/tip/ZhuLWZMLH18}
Z.~Zhu, J.~Lu, M.~Wang, S.~Zhang, R.~R. Martin, H.~Liu, and S.~Hu, ``A
  comparative study of algorithms for realtime panoramic video blending,''
  \emph{{IEEE} Trans. Image Processing}, vol.~27, no.~6, pp. 2952--2965, 2018.

\bibitem{conf/mm/MengWL15}
X.~Meng, W.~Wang, and B.~Leong, ``Skystitch: {A} cooperative multi-uav-based
  real-time video surveillance system with stitching,'' in \emph{Proceedings of
  the 23rd Annual {ACM} Conference on Multimedia Conference, {MM} '15,
  Brisbane, Australia, October 26 - 30, 2015}, 2015, pp. 261--270.

\bibitem{conf/icip/El-SabanEKR11}
M.~A. El{-}Saban, M.~Ezz, A.~Kaheel, and M.~Refaat, ``Improved optimal seam
  selection blending for fast video stitching of videos captured from freely
  moving devices,'' in \emph{18th {IEEE} International Conference on Image
  Processing, {ICIP} 2011, Brussels, Belgium, September 11-14, 2011}, 2011, pp.
  1481--1484.

\bibitem{journals/cgf/LinLCZ16}
K.~Lin, S.~Liu, L.~Cheong, and B.~Zeng, ``Seamless video stitching from
  hand-held camera inputs,'' \emph{Comput. Graph. Forum}, vol.~35, no.~2, pp.
  479--487, 2016.

\bibitem{journals/tip/GuoLHZZG16}
H.~Guo, S.~Liu, T.~He, S.~Zhu, B.~Zeng, and M.~Gabbouj, ``Joint video stitching
  and stabilization from moving cameras,'' \emph{{IEEE} Trans. Image
  Processing}, vol.~25, no.~11, pp. 5491--5503, 2016.

\bibitem{journals/tip/NieSZSL18}
Y.~Nie, T.~Su, Z.~Zhang, H.~Sun, and G.~Li, ``Dynamic video stitching via
  shakiness removing,'' \emph{{IEEE} Trans. Image Processing}, vol.~27, no.~1,
  pp. 164--178, 2018.

\bibitem{journals/pami/GioiJMR10}
R.~G. von Gioi, J.~Jakubowicz, J.~Morel, and G.~Randall, ``{LSD:} {A} fast line
  segment detector with a false detection control,'' \emph{{IEEE} Trans.
  Pattern Anal. Mach. Intell.}, vol.~32, no.~4, pp. 722--732, 2010.

\bibitem{journals/tog/LiuYT013}
S.~Liu, L.~Yuan, P.~Tan, and J.~Sun, ``Bundled camera paths for video
  stabilization,'' \emph{{ACM} Trans. Graph.}, vol.~32, no.~4, pp. 78:1--78:10,
  2013.

\bibitem{journals/tog/IgarashiMH05}
T.~Igarashi, T.~Moscovich, and J.~F. Hughes, ``As-rigid-as-possible shape
  manipulation,'' \emph{{ACM} Trans. Graph.}, vol.~24, no.~3, pp. 1134--1141,
  2005.

\bibitem{journals/jgtools/IgarashiI09}
T.~Igarashi and Y.~Igarashi, ``Implementing as-rigid-as-possible shape
  manipulation and surface flattening,'' \emph{J. Graphics, GPU, {\&} Game
  Tools}, vol.~14, no.~1, pp. 17--30, 2009.

\bibitem{journals/gandc/MartinezRF09}
F.~Mart{\'{\i}}nez, A.~J. Rueda, and F.~R. Feito, ``A new algorithm for
  computing boolean operations on polygons,'' \emph{Computers {\&}
  Geosciences}, vol.~35, no.~6, pp. 1177--1185, 2009.

\bibitem{journals/tog/HuangKAK14}
J.~Huang, S.~B. Kang, N.~Ahuja, and J.~Kopf, ``Image completion using planar
  structure guidance,'' \emph{{ACM} Trans. Graph.}, vol.~33, no.~4, pp.
  129:1--129:10, 2014.

\end{thebibliography}
\end{document}